\documentclass[twocolumn,showpacs,aps,prb]{revtex4}
 \usepackage{amsmath,amssymb,pstricks,pst-plot}

 \newcommand{\nts}[1]{\tmspace{-}{#1\thinmuskip}{#1\txtmu}}
\allowdisplaybreaks
\begin{document}

\title{Green's function of the half-filled Landau level Chern-Simons
  theory in the temporal gauge} 

\author{J. Dietel}
\altaffiliation{Current address:
Institut f\"ur Theoretische Physik, Freie Universit\"at Berlin, 
Arnimallee 14, D 14195 Berlin, Germany}
\affiliation{
Institut f\"ur Theoretische Physik, Universit\"at Leipzig,\\
Augustusplatz 10, D 04109 Leipzig, Germany \\
e-mail: dietel@itp.uni-leipzig.de\\}

\begin{abstract}
We study the Green's function of the $ \nu=1/2 $ Chern-Simons
system in the temporal
(Weyl) gauge. We derive the Chern-Simons path integral in the temporal
gauge. In order to do this, we gauge transform the path integral in the Coulomb gauge which
represents the partition function of the correct normal ordered Chern-Simons
Hamiltonian. We calculate the self energy of this path integral in
the random-phase approximation (RPA) for temperature $T=0 $. 
This self energy does not have the  divergence with the logarithm of
the area, which is known to imply 
the vanishing of the exact Green's function in the Coulomb gauge for
an infinite area. By Chern-Simons retransforming the path integral 
representing the Green's function in the temporal gauge we calculate
explicitly the exact Green's function under the neglection
of the interaction between the
electrons, getting a finite value. Furthermore, we give arguments that the
Green's function of the interacting system is also finite.
The non-vanishing of the Green's function for infinite area
is due to a dynamical creation
of the phase factors linking the created and annihilated particles with the 
particles in the ground state. The absence of these phase factors is the reason
for the vanishing of the Green's function in the Coulomb gauge.   
\end{abstract}

\pacs{71.10.Pm, 73.43.-f, 71.27.+a}

\maketitle

\section{Introduction}
The combination of an electronic interaction and a strong magnetic field
in a two--dimensional electron system yields a rich variety of phases.
These are best classified by the filling factor $\nu$, which is the electron
density divided by the density of a completely filled Landau level.
In the case of $\nu \cong 1/2$, the behaviour of the system resembles that of
a Fermi liquid in the absence of a magnetic field, or at small magnetic
fields. This effect can be explained with a new sort of quasi-particles:
at $\nu =1/2$, each electron combines with two flux quanta of the magnetic
field to form a composite fermion; these composite fermions then move
in an effective magnetic field which is zero on the average.
The interpretation of many experiments supports this picture.
We mention transport experiments with anti-- dots,
in which features of the resistivity are related to
closed loops of the composite fermions around the dots \cite{kan1},
and also focusing experiments \cite{sme1}. An overview of further
experiments can be found in \cite{wil1}. 
A field theoretical formulation of this composite fermion  picture was first
established by 
Halperin, Lee, and Read (HLR) \cite{hlr} as well as Kalmeyer and Zhang
\cite{kal1}.

HLR studied many physical quantities within the
random-phase approximation (RPA). Most prominent among these is the effective
mass of the composite fermions which is found to diverge at the Fermi
surface \cite{hlr,st1}. This is caused by the interaction of the composite
fermions via transversal gauge interactions. Later on, Shankar and
Murthy \cite{sh1} proposed a
new theory of the $ \nu=1/2 $ system. Based upon a transformation
of the Chern-Simons (CS) Hamiltonian one achieves a separation
of the magneto-plasmon oscillators from the total interaction of the
system. After restricting the number of the magneto-plasmon oscillators
to the number of electrons Shankar and Murthy obtain a finite
quasi-particle mass which scales with the inverse of the strength of the
Coulomb repulsion. In their derivation they calculated a smaller number of
self energy Feynman diagrams than in the RPA is calculated. Recently,
Stern et al. \cite{st2} calculated the self energy 
of the theory of Shankar and Murthy in RPA finding
the same divergence of the effective mass as HLR. Furthermore, they examined
the Lagrangian formalism in the temporal (Weyl) gauge of the CS
theory. This appears to be very similar to the Hamilton theory of Shankar and
Murthy \cite{mu1} which works with a Hilbert space consisting of the
electron plus the magneto plasmon degrees of freedom and an additional constraint on the Hilbert space.
In their paper, Stern et al. show that the quasi particles of both
theories contain  the dipole nature of
the $ \nu=1/2$ Rezayi-Read wave function \cite{re1} which has a
good overlap with the $ \nu=1/2 $ exact ground state for small systems.  
In this paper we consider the Lagrangian formulation of the
CS theory in the temporal gauge.
Up to  now it is not clear whether this theory and the Hamilton theory of Shankar and Murthy are in any relation. This is due to the difficulty in 
formulating a coherent state path integral for a Hamilton theory 
which contains a constraint including fermionic fields.   
We should mention that besides the theories of HLR and Shankar and Murthy
there are other alternative formulations of the CS theory
which appear to be  similar
to the CS theory of Shankar and Murthy \cite{pas1}.

As mentioned first in \cite{hlr} by semi-classical arguments
and showed further by us non-perturbatively  (calculated explicitly
for the non-interacting system) \cite{di2} the Green's function of the
$ \nu=1/2 $ CS system in the version of HLR (Coulomb gauge of the
Lagrangian formulation of the CS theory) vanishes
exponentially with an exponent proportional to $ \log(A) $ where $ A $ is the
area of the system. This is caused by the CS transformation
which effectively gives 
a velocity boost to every electron. This velocity boost results in a
one particle energy which diverges proportional to  $ \log(A) $.   
We further show in \cite{di2} that the $ A $ asymptotics of
the Hartree-Fock approximation of the Green's function is in accordance with
the exact Green's function (in contrast to the Green's function in
RPA) which is the theoretical justification for
formulating a perturbation theory around  the Hartree-Fock mean field.
This theory was discussed by us in \cite{di5}.
Due to this divergence it is difficult to formulate a
quasi-particle language 
for such a theory. When neglecting this $ \log(A) $ divergence in the
self energy one gets the diverging effective mass on the fermi-surface
discussed above.  
On the way to formulate a CS theory of the half-filled Landau level
with meaningful quasi-particles 
we consider in this paper the Green's function of the CS theory in the temporal gauge. This is the  Green's function for the new composite fermions. 
We will show that this Green's function does
not vanish with an exponent proportional to
$\log(A) $ (at least for the Coulomb interacting system). We will
calculate explicitly the Green's function in
position-time representation by neglecting the Coulomb interaction.
This yields a finite Green's
function. Furthermore, we show that
the Green's function should be also finite 
when taking into account the interaction between the electrons.
These are the main results of this paper.

To this end we derive at first by a gauge transformation of the
CS path integral in the Coulomb gauge \cite{di1} the CS
path integral in the temporal gauge. We have shown in \cite{we1} and
\cite{di1} that one is not allowed to carry  out the time slice continuum 
limit in CS path 
integrals of quantum Hall systems due to an
additional term in the path integral which supply the correct 
operator order  in the correspondent CS Hamiltonian.
When neglecting this additional term we get the familiar CS path
integral representing a gauge theory (e.g. \cite{ho1}). We will show 
that we get the same
result either by writing down this gauge path integral in the temporal gauge
or by gauge transforming the correct normal order CS path
integral in the Coulomb gauge. By calculating the self energy of this path
integral in RPA we get a singularity proportional to  $ 1/T $ 
where $ T $ is the temperature. This was calculated 
earlier by Stern et al. in \cite{st1}
for the Hamilton theory of Shankar and Murthy.

By retransforming the Green's function path integral to the electrons
we get an effective path integral action of a time dependent
Hamiltonian.
This time
dependent Hamiltonian describes electrons in a homogeneous magnetic field with
two separated magnetic strings of opposite strength at the positions of
the creation and the annihilation
operator of the Green's function, which are adiabatically switched on
until they get two flux quantas (for the $ \nu=1/2 $ system). 
By calculating the ground
state energy and the ground state wave function of this Hamiltonian
we derive explicitly for the non-interacting system a non-vanishing 
zero temperature Green's function.
We will show further that the reason for the
$ T \to 0 $ vanishing of the Green's
function in RPA is caused by getting a difference in the ground state
energy for the system taking into account the two strings in comparison
to the system without the strings. 
The RPA corresponds to the energy correction in second order
perturbation theory in the string strength.
Therefore, we calculate in this paper the energy difference
for the interacting electron system, getting a zero
energy difference. Thus, we see that the exact Green's function
in the temporal gauge should be also finite when taking into account
the Coulomb interaction between the electrons.
Furthermore, we will see that by switching on the magnetic strings the
creation and the annihilation operator in the Green's function
gets additional CS phases
linking them with all other  electrons in the ground state.
These phase factors do not exist  
in the comparable expression of the Green's function in the Coulomb
gauge. This is the reason for the vanishing of the Coulomb
gauged Green's function
with an exponent proportional to $ \log(A) $  
not existent in the temporal gauge  \cite{di2}.

The paper is organized as follows:\\
In section II we derive the CS path integral in the
temporal gauge from the path integral in the Coulomb gauge.
We compare the Green's functions in RPA of both gauges in section III.
In section IV we consider the Green's function in the temporal gauge
non-perturbatively.  

\section{The CS path integral in the Coulomb as well as the temporal gauge}
In this section 
we consider interacting spin polarized electrons moving in two dimensions in a
strong magnetic field $ B $ directed in the negative $ z $-direction.
The electronic density is chosen such that the lowest Landau
level of the  non-interacting system is filled to a fraction
$ \nu=1/ \tilde{\phi}$ where $ \tilde{\phi} $ is an even number. We are mainly
interested in $ \tilde{\phi}=2 $. The CS transformation is defined by 
\cite{zh1} 
\begin{equation} 
 \Psi^+(\vec{r})=\Psi^+_{e}(\vec{r})
 \exp\left[i \tilde{\phi}\int d^2 r' \mbox{arg}(\vec{r}-\vec{r}\,') \,
 \rho(\vec{r}\,')\right] \,.  \label{5} 
\end{equation}
where $\Psi^+_{e}(\vec{r}) $ is the electron creation operator,
$\Psi^+(\vec{r})$ is the creation operator of the transformed fermions
(composite fermions),
$ \hat{\rho}(\vec{r}) $ is the density operator of the fermion
operators , and
$ \mbox{arg}(\vec{r}) $  is the angle that $ \vec{r} $ forms with the
$ x $-axis. In this paper we use the convention that
$ \mbox{arg} $ has its cut on the negative real axis.  
The Hamiltonian is given after the
transformation as:
\begin{align}
&  H_{CS}(\vec{a}_{CS})  = \int d^2r
\Bigg\{\frac{1}{2m}\Psi^+(\vec{r}) \big(-i\vec{\nabla}+\vec{A}  
+ \vec{a}_{CS}\big)^2 \Psi(\vec{r}) \nonumber   \\
&  \nts{2}   +\nts{1} \frac{1}{2} \nts{1} \int d^2r' 
 (|\Psi(\vec{r})|^2-\rho_B)
 V^{ee}_{\vec{r},\vec{r}\,'}(|\Psi(\vec{r}\,')|^2-\rho_B)\Bigg\}.
  \label{10}
\end{align}
The CS vector potential $ \vec{a}_{CS} $ is defined
by $ \vec{a}_{CS}(\vec{r})=\tilde{\phi} \int d^2r' \;\vec{f}(\vec{r}-\vec{r}\,')
\Psi^+(\vec{r}\,')\Psi(\vec{r}\,') $. 
Here $ \Psi^+(\vec{r}) $ creates (and $ \Psi(\vec{r}) $ annihilates) a
composite fermion
with coordinate $ \vec{r} $. $ V^{ee}_{\vec{r},\vec{r}\,'}=
e^2/|\vec{r}-\vec{r}\,'| $ is the Coulomb interaction
where $ e^2=q_e^2 /\epsilon  $. $ q_e $ is the charge of the electrons and
$ \epsilon $ is the dielectric constant of the background field $ \rho_B $.
$ \vec{A}(\vec{r}) $ is the vector potential 
$ \vec{A}=\vec{B} \times \vec{r} /2$ and $\vec{B} $ is a 
homogeneous magnetic field in the negative z-direction
$ \vec{B}=-B \vec{e}_z $ where $ \vec{e}_z $ is the
unit vector in $z $-direction. 
We suppose throughout this paper that $ B $ is a positive number. 
The function $ \vec{f}(\vec{r}) $ is given by 
$ \vec{f}(\vec{r})=\vec{e}_z \times \vec{r}/r^2 $.
We used 
the convention $ \hbar=1 $ and $ c=1$ in the above formula
(\ref{10}). Furthermore, we set $ q_e=1 $ for the coupling of the magnetic
potential to the electrons.
We obtained in \cite{di1} the partition function of
the Hamiltonian (\ref{10}) in the path integral formalism.
With the help of the real bosonic CS fields ($ a^0(r,t), \vec{a}(r,t) $) 
we get  
\begin{eqnarray}
Z_{\mbox{\scriptsize Coul}}& = &  \lim_{\epsilon \to 0} \frac{1}{\overline{N}}
\prod\limits_{l=1}^{N_l}
\int_{ \mbox{\scriptsize BC}} {\cal D}
[a^0_l,\vec{a}_l]\, {\cal D}[\Psi^*_l,\Psi_l] \label{20} \\
& & \times 
\exp\left[-\epsilon\left( 
L^{\mbox{ \scriptsize Coul}}_l+L_{CS,l}
+L_{ee,l}\right)\right] \nonumber \;. 
\end{eqnarray}
The various functions in (\ref{20}) are given by 
\begin{eqnarray}
  L^{\mbox{ \scriptsize Coul}}_l &  = &      \int d^2r\; \Psi^*_{l}(\vec{r})
 \frac{1}{\epsilon} \left(\Psi_l(\vec{r})- \Psi_{l-1}(\vec{r})\right)
 \label{30}\\
 & &    -\Psi^*_{l}(\vec{r}) \left( \mu+\left(1+i
    \frac{\epsilon}{2}a^0_{l}(\vec{r})\right)ia^0_{l}(\vec{r})\right)
 \Psi_{l-1}(\vec{r})\nonumber\\
& &     + \frac{1}{2m}\Psi^*_{l} (\vec{r})
\left(-i\vec{\nabla}+\vec{A}(\vec{r})+\vec{a}_l(\vec{r})
\right)^2\Psi_{l-1}(\vec{r}) \;,\nonumber  \\
 L_{CS,l} &  = &     
\frac{1}{4\pi\tilde{\phi}}\int
d^2r \, i  a^0_{l}(\vec{r})\, \vec{\nabla} \times \,\vec{a}_{l}(\vec{r}) 
\label{40} \;,\\  
 L_{ee,l} & = &    
\frac{1}{2} \int d^2rd^2r' \,
  (\Psi^*_{l}(\vec{r})\Psi_{l-1}(\vec{r})-\rho_B)  \nonumber  \\   
 & &\qquad  \times  V^{ee}_{\vec{r},\vec{r}\,'}\;
 (\Psi^*_{l}(\vec{r}\,')\Psi_{l-1}(\vec{r}\,')-\rho_B) 
  \label{50} 
\end{eqnarray}
and
\begin{equation}
\overline{N}=\prod\limits_{l=1}^{N_l}\int_{ \mbox{\scriptsize BC}}{\cal D}[a^0_l,\vec{a}_l]
\exp\left[-\epsilon\left(L_{CS,l}\right)\right]      \;.
\label{60}
\end{equation}
The path integral (\ref{20}) is correct under the  
gauge condition $ \vec{\nabla} \cdot \vec{a}_l=0 $ (Coulomb gauge). 
The time slice width $ \epsilon $ is defined by  $\epsilon=\beta/N_l $
where $ \beta=1/T $. 
The index $l $ counts the discrete time slices. 
Furthermore, we have anti-periodic boundary conditions $\Psi_{N_l}=-\Psi_{0}$
(denoted by BC) for the Grassmann fields.
The action of the path integral (\ref{20}) is given by a fermionic term $
L^{\mbox{ \scriptsize Coul}}_{l} $, a bosonic term
$ L_{CS,l} $ of the CS form, and a 
Coulomb interaction term $ L_{ee,l}$. 
In
comparison to the CS path integral of HLR \cite{hlr} 
we get an additional term proportional 
to $ \epsilon (a^0_l)^2\Psi^*_{l}\Psi_{l-1}/2$  in
$ L_l^{\mbox{ \scriptsize Coul}} $
(\ref{30}). 
This term is due to the non-normal-order 
of the $ \Psi^6 $ term in the CS Hamiltonian $ H_{CS} $ 
(\ref{10}). This is best seen by integrating the path integral 
(\ref{20}) over the CS fields. Due to the additive term 
one can not perform the formal limit $\epsilon \to 0 $ in (\ref{20}). \\
Now suppose that one may neglect the
$ \epsilon (a^0_l)^2\Psi^*_{l}\Psi_{l-1}/2$
term  in $ L^{\mbox{ \scriptsize Coul}}_l $
(\ref{30}). Then one can  take
the formal limit $\epsilon \to 0 $ in (\ref{20}) getting the well known 
path integral describing a CS gauge theory in the Coulomb gauge.
Without the gauge fixing condition $ \vec{\nabla} \cdot \vec{a}_l(\vec{r})=0 $
the path integral consists of the three
independent CS fields $ a^0 $ and $ \vec{a} $ \cite{ho1}.
The CS theory in the temporal 
gauge \cite{sh1,st2} is then given by $ a^0=0 $.
The neglection of  the term $\epsilon
(a^0_l)^2\Psi^*_{l}\Psi_{l-1}/2 $
in $ L^{\mbox{ \scriptsize Coul}}_l $ is to our opinion not satisfactory
because we showed in \cite{we1,di1} that this results 
in the wrong RPA energy. Thus, it is important
to determine whether the CS path integral in the temporal gauge
used by Shankar and Murthy \cite{sh1} and Stern et al. \cite{st2}
is correct by considering
the full Lagrangian (\ref{30}) in the derivation. This will be done in the
following:

We start from the path integral (\ref{20}) by the 
gauge transformation of the fermionic fields:
\begin{eqnarray}
  \Psi_l & \rightarrow & \exp\left[i \left( g_l+
    F(a^0,\vec{a})\right)\right] \Psi_l \,, \nonumber \\        
  \Psi_0 & \rightarrow &  \exp\left[i  F(a^0,\vec{a})\right] \Psi_0                               \label{100} 
\end{eqnarray}
with
\begin{equation}
  g_l=\sum\limits_{k=1}^l
    \epsilon \left(a^0_k - \frac{1}{\beta}\sum\limits_{k=1}^{N_l}
      \epsilon a^0_k\right)
    \,.      \label{110} 
\end{equation}
The definition of $ g_l $ is chosen such that the Fourier
transformation of $ g_l $ is $ 1/(i \omega) $ times the Fourier
transformation of $ a^0 $ (for $ \epsilon \to 0 $). 
$ F(a^0,\vec{a}) $ is a function of the fields $ a^0 $ and
$ \vec{a} $ which does not depend explicitly on the time index $ l $.
One gets from this transformation that the new Grassmann
fields keep the anti-periodic boundary condition $\Psi_{N_l}=-\Psi_{0}$.
In the following, we define $ F(a^0,\vec{a}) $ such that 
the ($ \omega=0 $)-term of the  Fourier transformation of the function   
$ g_l+F(a^0,\vec{a}) $ is zero.
This results in
\begin{equation}
F(a^0,\vec{a})=-\left( \frac{1}{\beta} \sum\limits_{k'=1}^{N_l}
  \epsilon \sum\limits_{k=1}^{k'} \epsilon a_k^0-\frac{1}{2}
  \sum\limits_{k=1}^{N_l} \epsilon a_k^0 \right) \,. \label{120}
\end{equation}
After inserting the transformation (\ref{100}) of the Grassmann fields
in (\ref{20})
we expand the exponential function in (\ref{100}).
We do not have to consider all of the expansion terms 
for $ \epsilon \to 0 $. In order to determine which expansion terms
have to be considered we further expand the exponential function
in (\ref{20})
of the exponent   $ L^{\mbox{ \scriptsize Coul}}_l $ containing at least one
CS field $ a^0_l $ or $ \vec{a}_l $. Now one may assume that it is enough to
consider only linear terms in the expansion of the exponential function in
(\ref{100}). This is not correct because one gets also terms of the order
$ O(1) $ after integrating out the Chern-Simons fields (e.g. a sum of
Grassmann fields over the time slices times $ \epsilon $ is of order
$ O(1) $). By analyzing the terms carefully we see that one
has to take into
account  up to the quadratic
expansion terms in the exponential function in (\ref{100}) to get all
$ O(1) $ terms in the path integral. 
Doing so, one can observe the interesting effect that the 
$\epsilon (a^0_l)^2\Psi^*_{l}\Psi_{l-1}/2 $ term in (\ref{30}) is
cancelled by some of the expansion terms in (\ref{100}).
After an additional gauge transformation 
\begin{eqnarray}
  \Psi_l(\vec{r}) & \rightarrow & \exp\left[i \tilde{\phi} \vec{f}(0)
  \vec{r} \right] \Psi_l(\vec{r})                \label{130}
\end{eqnarray}
we obtain a path integral in which it is allowed to take the limit  $ \epsilon \to 0 $.
This path integral is given by
\begin{eqnarray}
Z_{\mbox{\scriptsize Weyl}}& = & \frac{1}{\overline{N}} 
\int_{\mbox{\scriptsize BC}} {\cal D}[a^0,\vec{a}]
\, {\cal D}[\Psi^*,\Psi] \label{140} \\
& & \times 
\exp\left[- \int_0^{\beta} dt \left( 
L^{\mbox{\scriptsize Weyl}}+L_{CS}+
L_{ee}\right)\right] \nonumber 
\end{eqnarray}
with 
\begin{align}
&   L^{\mbox{\scriptsize Weyl}}  \nts{2} =  \nts{2}     \int \nts{2} d^2r \, \Psi^*(\vec{r},t)
  \bigg( \nts{1}\partial_t-\mu- \frac{i}{\beta}  
  \int_0^{\beta} dt' a^0(\vec{r},t')\bigg) \Psi(\vec{r},t) \nonumber  \\
 &  \qquad  + \frac{1}{2m}\Psi^*(\vec{r},t)
\bigg (-i\vec{\nabla}+\vec{A}(\vec{r}) 
+\vec{a}(\vec{r},t) \nonumber  \\
& \qquad \qquad +\vec{\nabla}
 \left(g(\vec{r},t)+F(a^0,\vec{a})\right) \bigg)^2\Psi(\vec{r},t)
  \;,\label{150}   \\
&  L_{CS} =   \frac{1}{4\pi\tilde{\phi}}\int
d^2r\; \; i  a^0(\vec{r},t)\, \vec{\nabla} \times \,\vec{a}(\vec{r},t) 
\label{160} \;,\\  
&  L_{ee}   =      
\frac{1}{2} \int d^2rd^2r' \,
  (\Psi^*_{l}(\vec{r})\Psi_{l-1}(\vec{r})-\rho_B)  \nonumber  \\   
 & \qquad \qquad \qquad  \times  V^{ee}_{\vec{r},\vec{r}\,'}\;
 (\Psi^*_{l}(\vec{r}\,')\Psi_{l-1}(\vec{r}\,')-\rho_B) \, .   \label{170}      
\end{align}
By the neglection of  the third term in the first bracket in
$ L^{\mbox{\scriptsize
    Weyl}} $ for $ T=0 $
and the definition of the longitudinal CS gauge potential
\begin{equation}
  \vec{a}_L(\vec{r})=\vec{\nabla}
 \left(g(\vec{r})+F(a^0,\vec{a}) \right) \label{180}
\end{equation}
we get the well known CS path integral in the temporal
gauge \cite{sh1,st2}. 
This path integral was used by Stern et al. in  \cite{st2} to show that
the quasi-particles
in the temporal gauge behave like dipoles with a dipole momentum perpendicular
to their canonical momentum (for small momentum and frequency). This
can be seen by calculating
the response of the electrons in the RPA due to some external
potential. This picture of the CS quasi-particles is very attractive
\cite{si1} due to
a similar dipole interpretation of the Rezayi-Read wave function \cite{re1}.
It has been shown that this
wave function has a very good overlap with the exact ground state for small
systems \cite{re1}.

\section{The RPA Green's functions}
In this section we determine the RPA Green's functions in the Coulomb as well
as the temporal gauge for temperature $ T \to 0 $. This
was done earlier for the Coulomb gauge \cite{di2}.
Since the Coulomb interaction has no influence on the singularity
of the Green's function in the Coulomb as well as in the temporal gauge
we will simplify the notation by considering explicitly only the interaction 
free case of the Green's function. 
The Coulomb interaction can easily be taken into account by carrying out 
a Hubbard-Stratonovich decoupling \cite{ne1} of the Coulomb
interaction term (\ref{170}).
In the following, we will mention explicitly where the results for the non-interacting system differ from those of the interacting system.

In \cite{di1}, we calculated
the grand canonical
potential $ \Omega_{\mbox{\scriptsize Coul}} $ from the CS path integral
in the Coulomb gauge (\ref{20}) 
in RPA. This was done by carrying out the integration of (\ref{20}) over the
fermionic fields and further by expanding the logarithms of the result
in the CS fields. The restriction to quadratic order in the RPA for the 
CS fields results in 
\begin{equation}
\Omega^{\mbox{\scriptsize Coul}}=\frac{1}{2 \beta }\sum\limits_{\vec{q}, \Omega}
\log\left(1-\Pi_{00}
    (\Pi_{TT}+\frac{\rho}{m})\frac{(2 \pi \tilde{\phi})^2}{q^2}\right)\,.\label{190} 
\end{equation}  
In this equation $ \Pi_{00} $ is the ideal gas density-density response
and $\Pi_{TT} $ is the transversal momentum-momentum response.
These response functions can be calculated exactly \cite{di1}. 
The grand canonical potential is then a functional of the interaction free
Green's function $ G=-1/(i\omega-q^2/(2m)+\mu) $.   
The RPA
self energy can be calculated by  $ \Sigma^{ \mbox{\scriptsize Coul}}=
\delta \Omega^{\mbox{\scriptsize Coul}}/\delta G $.
By carrying out the calculation for the path integral (\ref{20}), one gets for 
$ \Sigma^{\mbox{\scriptsize Coul}}$ one
term which is divergent for $ A \to \infty $ \cite{di2}. This term
corresponds  to a self energy Feynman-diagram with one density-density
$ (a^0, a^0) $ RPA vertex. The other RPA self energy diagram containing
one transversal momentum-momentum $ (\vec{a}_T, \vec{a}_T) $ vertex is finite
(here $ \vec{a}_T $ is the transversal component of $ \vec{a} $).
One obtains for the divergent self-energy term  
\begin{align} 
& \Sigma^{ \mbox{\scriptsize Coul}}_{0
    0}(k,\omega)=\Sigma_F(k,\omega) \label{200} \\
& 
  + \frac{1}{\beta} \sum\limits_{\vec{q}, \Omega} G(\vec{k}+\vec{q},
\omega+\Omega) \left({\cal D}_{00}(q,\Omega)-\frac{\rho}{m}\right)
 \frac{(2 \pi \tilde{\phi})^2}{q^2}   \nonumber 
\end{align}
 with the Fock-self energy 
\begin{equation} 
\Sigma_{F}(k,\omega)= - \sum\limits_{\vec{q}} n_F(|\vec{k}+\vec{q}|)
\frac{\rho}{m} \frac{(2 \pi \tilde{\phi})^2}{q^2}  \label{203}
\end{equation}
and the $ (a^0, a^0) $ RPA-vertex 
\begin{equation}
 {\cal D}_{00}(q,\Omega)=
 \frac{(2 \pi \tilde{\phi})^2}{q^2}\frac{(\Pi_{TT}+\frac{\rho}{m})}{1-\Pi_{00}
    (\Pi_{TT}+\frac{\rho}{m})\frac{(2 \pi \tilde{\phi})^2}{q^2}}\,. \label{210}
 \end{equation}
$ \rho $ is the density of the system. 
The singular part of the $ (a^0,a^0) $ RPA vertex $ {\cal D}_{00} $ in
(\ref{200}) has 
its parameter range in $ \Omega \gg q \sqrt{\mu/m} $ and $ q^2/(m\mu) \ll 1$ 
.\cite{hlr} In this range one gets for the vertex  
\begin{equation}
 {\cal D}_{00}(q,\Omega)\approx
 \frac{(2  \pi \tilde{\phi})^2 }{q^2} \frac{\rho}{m}
 \frac{\Omega^2}{\Omega^2+\omega_c^2}
 \label{206}
\end{equation}
Here $ \omega_c $ is given by $ B/m=(2 \pi \tilde{\phi}) \rho/m  $.
With the help of this expression we obtain for the $ A \to \infty $
singular part of the
Green's function 
\begin{align} 
&\Sigma^{ \mbox{\scriptsize Coul}}_{00}(k,\omega)=  \label{220}\\
&  \tilde{\phi}^2 \frac{\mu}{2}
\log\left(\frac{1}{A}\right)
\frac {\left(i \omega-\frac{k^2}{2m}+\mu\right)}{\left(\omega_c+
\mbox{sgn}\left[\frac{k^2}{2m}-\mu\right]\left(\frac{k^2}{2m}-i \omega-\mu
\right)\right)} \,. \nonumber 
\end{align}
$ \mbox{sgn}[\cdot] $ is the sign of the
argument.

We may now carry out a similar calculation for the CS path integral in the
temporal gauge (\ref{140})
yielding for the grand canonical potential
$ \Omega^{\mbox{\scriptsize Weyl}} $ in RPA 
\begin{align}
& \Omega^{\mbox{\scriptsize Weyl}} =  \frac{1}{2 \beta}
\nts{2} \sum\limits_{\vec{q}, \Omega \not=0} 
\nts{2} \log \nts{1} \left( \nts{1} 1-(\Pi_{LL}+\frac{\rho}{m})
  (\Pi_{TT}+\frac{\rho}{m})
    \frac{(2 \pi \tilde{\phi})^2}{\Omega^2} \nts{1} \right) \nonumber \\
  &  +\frac{1}{2\beta}\sum\limits_{\vec{q}}
\log\left(1-\Pi_{00}(\vec{q},0) 
    (\Pi_{TT}(\vec{q},0)+\frac{\rho}{m})\frac{(2 \pi \tilde{\phi})^2}{q^2}\right)
.  \label{230} 
\end{align}  
Here $ \Pi_{LL} $ is the ideal gas longitudinal momentum-momentum response
function. With the help of the ideal gas continuity equation
$ (\Pi_{LL}+\rho/m)/\Omega^2=\Pi_{00}/q^2 $ we get
$ \Omega^{\mbox{\scriptsize Coul}}=\Omega^{\mbox{\scriptsize Weyl}}$.
Nevertheless, $ \Omega^{\mbox{\scriptsize Coul}} $ and
$ \Omega^{\mbox{\scriptsize Weyl}}$ are not identical as a function
of $ G $. As in the Coulomb gauge we calculate the RPA self energy by  
$  \Sigma^{ \mbox{\scriptsize Weyl}}=\delta \Omega^{\mbox{\scriptsize
    Weyl}}/\delta G $.
Then we get one term which corresponds to the  RPA self energy diagram
containing one transversal momentum-momentum vertex
$ (\vec{a}_T, \vec{a}_T) $.
This finite term is the same as in the Coulomb gauge. 
The other term corresponds to the self energy diagrams containing one
longitudinal momentum-momentum vertex $ (\vec{a}_L, \vec{a}_L) $.
It is given by
\begin{align} 
& \Sigma^{ \mbox{\scriptsize Weyl}}_{LL}(k,\omega)= \frac{1}{\beta}  \sum\limits_{\vec{q}}  G(\vec{k}+\vec{q},
\omega) {\cal D}_{00}(q,0) \label{240} \\
& +\frac{1}{\beta}  \sum\limits_{\vec{q},\Omega \not=0 }  G(\vec{k}+\vec{q},
\omega+\Omega) {\cal D}_{00}(q,\Omega)
 \frac{q^2}{\Omega^2}
 \left( \frac {(2 \vec{k}+\vec{q}) \vec{q}}{2m q}\right)^2 .  \nonumber  
\end{align}
Here the  first term originates from the third term in the first bracket in
(\ref{150}).
Contrary to the Coulomb gauge, one finds an infinite 
self energy for $ T=0$ in the parameter range 
$ \Omega \ll q \sqrt{\mu/m} $ and $ q^2/(m\mu) \ll 1 $
because of the additional 
$ 1/\Omega^2 $ factor in the second term in (\ref{240}).
In this parameter range the vertex $ {\cal D}_{00} $ is given by  
\begin{eqnarray}
& &  {\cal D}_{00}(q,\Omega)  \approx \label{250} \\
& &  \frac{q^2}{24 \pi m} \frac{1}{\sqrt{2 m \mu}\frac{m |\Omega|}{(2 \pi)^2q}
   +\frac{e^2
 m}{(2 \pi \tilde{\phi})^2}q+q^2 \left(\frac{1}{48 \pi^2 }+\frac{1}{(2\pi \tilde{\phi})^2}\right)} \,.
 \nonumber 
\end{eqnarray}
This expression contains a correction due to the Coulomb
interaction. 
Inserting (\ref{250}) in (\ref{240}) results in a self energy term
proportional to $ \beta $ which is given by
\begin{align}
\Sigma^{ \mbox{\scriptsize Weyl}}_{LL}(k,\omega) \approx
\frac{\beta}{12}  \sum\limits_{
|q|<\kappa}
G(\vec{k}+\vec{q},\omega)
{\cal D}_{00}(q,0)
\left( \frac {(2 \vec{k}+\vec{q}) \vec{q}}{2m}\right)^2 \nts{3}. \label{260} 
\end{align}
Here $ \kappa \ll \sqrt{m \mu}$ is a momentum cut off. Thus, one finds that 
$ \Sigma^{ \mbox{\scriptsize Weyl}}_{LL} $ is proportional to $ \beta $ and 
diverges as $ T \to 0$. 
This results in a divergent self energy for temperature $ T=0 $.
By comparing (\ref{260}) with (\ref{220}) one sees that
the self energy in the temporal gauge is not divergent  for $A \to \infty $.
In \cite{di2} we showed that the Green's function in the Coulomb gauge
vanishes with an exponent
proportional to $ \log(A) $ in the position-time representation.
This is caused by an effective velocity boost obtained for every CS
quasi-particle by the CS transformation \cite{hlr}. This results in a self
energy proportional to $ \log(A) $. Due to the missing
$ \log(A) $ term in the RPA self energy in the temporal gauge
the first question we want to answer 
in the following subsections by non-perturbative methods is
\begin{itemize}
\item[{\bf 1.}]Is it true that the exact CS Green's function in the
  temporal gauge does not show a similar $ A \to \infty $ vanishing asymptotics
   as the Green's function in the Coulomb gauge? 
\end{itemize}

The $ \beta $-divergence of the self energy in
(\ref{260}) has its origin in the form of the $ \vec{a}_L $ coupling to
the fermionic fields in 
$ L^{\mbox{\scriptsize  Weyl}} $  (\ref{150})
which results in an additional $ q^2/\omega^2 $ factor for every
 $ (\vec{a}_L, \vec{a}_L) $ vertex in comparison to the
 $ (a_0,a_0) $ vertices in the
 Coulomb gauge. From this it is clear that a similar kind of divergence
should also be given in the self energy diagrams beyond RPA.

Stern et al. mentioned first in \cite{st2}
(for the case of the Hamiltonian formulation of the
CS theory in the temporal gauge) that this divergence in the self energy is
caused by the additional gauge freedom of the CS path integral 
(\ref{140}) with respect to time independent gauge
transformations. This causes the
partition function to be independent of the zero frequency $ \vec{a}_L$
variable by additionally carrying out the integration over 
the fermionic fields.
Nevertheless, one can not
deduce from this the behaviour of the partition function for $ \vec{a}_L$
with frequencies approximately zero by additionally
carrying out the integration over the fast
$ \vec{a}_L$ modes and $ \vec{a}_T $. This would be 
more relevant for the
behaviour of the Green's function than the additional gauge freedom.  
Thus, we are led to the second  question we would like to answer 
in the following subsections by non-perturbative methods 
\begin{itemize}
\item[{\bf 2.}] Is it true that the exact CS Green's function in the
  temporal gauge is zero for temperature $ T=0 $? 
\end{itemize}

A first approach to a solution of this question is given by the
following observation:
By carrying out a gauge retransformation (\ref{100}) of the path integral
representing the Green's function $ G( \vec{r} ,t) $
in the temporal gauge the Grassmann fields 
representing the created  and the annihilated particle in the Green's function
get exponential prefactors of the form  (\ref{100}).
By an expansion of the exponents up to quadratic order in the $ a^0 $ fields  
we get for the $ T \to 0 $ diverging part  of the
Green's function a term proportional to $ \beta ({\cal D}^{\nts{2} \mbox{ \scriptsize ex}}_{00}(\vec{r} ,\omega=0)-
{\cal D}^{\nts{2}\mbox{ \scriptsize ex}}_{00}(0,\omega=0)) $. 
$ {\cal D}^{\nts{2}\mbox{ \scriptsize ex}}_{00}(\vec{r} ,\omega=0) $ is the exact
$ (a^0, a^0) $ vertex corresponding to (\ref{210}) in RPA.
$ {\cal D}^{\nts{2} \mbox{ \scriptsize ex}}_{00}(\vec{r} ,\omega=0) $ is given by
\begin{align}
&  {\cal D}^{\nts{2} \mbox{ \scriptsize ex}}_{00}(\vec{r} ,\omega=0)= -
\tilde{\phi}^2
\int   d^2r' \vec{f}(\vec{r}-\vec{r}\,')\cdot \vec{f}(\vec{r}\,')
\frac{\langle \, \hat{\rho}(\vec{r}\,'  )\rangle}{m} 
\label{270} \\
& + \tilde{\phi}^2 \nts{2} \int \nts{2} dt \nts{2} \int \nts{2}  d^2r' d^2r''\nts{1}
\vec{f}(\vec{r}-\vec{r}\,')\cdot  \langle
T \hat{\vec{J}}(\vec{r}\,',t´)  \hat{\vec{J}}(\vec{r}\,'',0 ) \rangle_c
\cdot \vec{f}(\vec{r}\,'').    \nonumber 
\end{align}
$ T $ is the time ordering operator.
The right hand side of (\ref{270}) can be calculated in the
electronic system.
Thus,  $ \hat{\rho}(\vec{r}) $ is the density operator and
$ \hat{\vec{J}}(\vec{r}) $ is the current
operator of the electrons. $ \langle \cdot \rangle_c $ is the connected average
with respect to the electronic ground state (not the CS ground state).
It is clear from the derivation above that the Feynman diagrams
of the Green's function
in first order in $ \Sigma_{LL}^{\mbox{\scriptsize Weyl}} $
are contained in this Green's function. 
We now restrict our considerations to the non-interacting electron system. 
In appendix A
we calculate $ {\cal D}^{\mbox{\scriptsize ex}}_{00}(\vec{r} ,\omega=0) $
without any
approximation for this non-interacting system.
We obtain 
\begin{equation} 
{\cal D}^{\mbox{\scriptsize ex}}_{00}(\vec{r} ,0)=
\frac{\tilde{\phi}}{l_0^2 m}\left[ \nts{1}
\left( \nts{1} \log\left(\frac{r^2}{2 l_0^2}\right)+\gamma \nts{1} \right)
\nts{1} e^{-\frac{r^2}{2 l_0^2}}
+\frac{3}{2}  E_{1} \nts{1} \left(\nts{2} \frac{r^2}{2 l_0^2} \nts{2} \right)
\right] \nts{1} . \label{280}
\end{equation}
Here $ l_0=1/\sqrt{B} $ is the magnetic length. 
$ \gamma $ is Euler's constant and $ E_1 $ is the exponential
integral function. 
We obtain that
$ {\cal D}^{\nts{2}\mbox{ \scriptsize ex}}_{00}(\vec{r} ,\omega=0)-{\cal D}^{\mbox{ \scriptsize ex}}_{00}(0 ,\omega=0) $
diverges for the non-interacting system. This divergence was
regularized in 
the RPA self energy formula (\ref{240}) with  the help of a
momentum cut-off in the UV region.
It is easy to see that the expression   
$  {\cal D}^{\nts{2}\mbox{ \scriptsize ex}}_{00}(\vec{r} ,\omega=0)-{\cal
  D}^{\nts{2}\mbox{ \scriptsize ex}}_{00}(0 ,\omega=0) $ agrees 
with the energy formula of the  second order perturbation theory
of electrons in a
homogeneous  magnetic field $ B $ under the perturbation of two magnetic strings
of flux $ \tilde{\phi} $ and $-\tilde{\phi} $ at the origin and at the position
$ \vec{r} $, respectively (see e.g. appendix A).
It is well known that the energy 
of electrons in a homogeneous 
magnetic field $ B $ with two magnetic strings is finite. Therefore, 
this energy can not be calculated perturbatively. We will show
in the following subsections that this is in fact the reason for the
temporal gauged Green's function in RPA to be zero for $ T=0 $. Furthermore,
we show that the exact Green's
function is finite for $ T=0 $ because the ground state energy corrections
due to the two magnetic strings is zero.

\section{The exact CS Green's function in the temporal gauge}
In this section, we calculate the CS Green's function in the
temporal gauge non-perturbatively. This was done by us in \cite{di3} for the
CS Green's function in the Coulomb gauge. There we
determined the Green's function in the position-time representation by
CS retransforming (\ref{5}) the Green's function to the electronic
Hilbert space.
A similar procedure will be done in this section for the Green's function
in the temporal gauge. It is defined by
\begin{equation}
 G^{\mbox{\scriptsize{Weyl}}}(\vec{x},\vec{x}\,';\tau,\tau')=
   \langle \Psi(\vec{x},\tau) \Psi^*(\vec{x}\,',\tau')
   \rangle_{\mbox{\scriptsize{Weyl}}} \,. 
   \label{290}
\end{equation}
Here $ \langle \cdot \rangle_{\mbox{\scriptsize{Weyl}}} $ is the average
with respect to the CS path integral in the temporal  gauge (\ref{140}).
We now carry out the inverse of the gauge transformation (\ref{100})
and (\ref{130})
on the fermionic fields in this expression. 
After  integrating  out the CS fields one gets
the following expression
\begin{align}
& G^{\mbox{\scriptsize{Weyl}}}(\vec{x},\vec{x}\,';\tau,\tau') = \frac{1}{N_G}
 \label{300} \\
& \nts{2} \times \nts{2} \int_{ \mbox{\scriptsize BC}} \nts{1}
 {\cal D}[\Psi^*,\Psi] 
\Psi(\vec{x},\tau) \Psi^*(\vec{x}\,',\tau') \exp\left[- \int_0^{\beta} dt  
L_G(\vec{A}^t_{ss})\right]  . \nonumber 
\end{align}
with the norm $ N_G $
\begin{equation}
N_G = \int_{\mbox{\scriptsize BC}}
 {\cal D}[\Psi^*,\Psi] \exp\left[- \int_0^{\beta} dt  
L_G(0)\right]  \label{310}  
\end{equation}
and the Lagrangian
\begin{align}
& L_G(\vec{A}^t_{ss}) =       \int d^2r \Bigg\{ \Psi^*(\vec{r},t)
  \bigg( \partial_t-\mu\bigg) \Psi(\vec{r},t) \label{320} \\
&+ \frac{1}{2m}\tilde{\phi}^2 \nts{2} \int d^2 r' \vec{f}^2(\vec{r}-\vec{r}\,')
\Psi^*(\vec{r},t)\Psi(\vec{r},t) \Psi^*(\vec{r}\,',t)\Psi(\vec{r}\,',t)
\nonumber  \\
&  +\nts{1} \frac{1}{2m}\Psi^*(\vec{r},t)
\nts{1}\bigg(\nts{3} -i\nts{1}\vec{\nabla}+\vec{A}(\vec{r}) 
+\vec{a}_{CS}
(\vec{r},t)+\vec{A}^t_{ss}(\vec{r})\bigg)^2 \nts{3} \Psi(\vec{r},t)
\nonumber \\ 
& + \frac{1}{2} \nts{1} \int d^2r' 
 (|\Psi(\vec{r},t)|^2-\rho_B)
 V^{ee}_{\vec{r},\vec{r}'}(|\Psi(\vec{r}\,',t)|^2-\rho_B)\Bigg\}. \nonumber 
\end{align}
The string configuration $ \vec{A}^t_{ss} $  in $L_G(\vec{A}^t_{ss})$ is given by  
 \begin{eqnarray}
 \vec{A}^t_{ss}(\vec{r}) & = & -\tilde{\phi}\vec{f}(\vec{r}-\vec{x})
  \left( \Theta(\tau-t)-\frac{\tau-t}{\beta}-\frac{1}{2} \right) \label{330}
   \\
 & & +\tilde{\phi}\vec{f}(\vec{r}-\vec{x}\,')\left( \Theta(\tau'-t)
  -\frac{\tau'-t}{\beta}-\frac{1}{2} \right) \nonumber \,.
 \end{eqnarray}
 Here $\Theta(x) $ is the Heavyside function. 
We see from this formula that the effective Lagrangian of the Green's
function is given by
the well known CS Lagrangian and additionally two time dependent
strings with opposite fluxes  centered at the coordinates $ \vec{x} $ and
$ \vec{x}\,'$. Furthermore, we see from (\ref{330}) that the Green's function
depends on $ \tau-\tau'$  (time translational invariance).
We see from (\ref{300})-(\ref{330}) that one has to solve a complicated
time dependent Schr\"odinger equation to get the Green's function in the
 temporal gauge. Nevertheless, we will derive a solution of the problem
 for temperature $ T=0 $ below. Thus, in the following we restrict the
 calculation of the Green's function to temperature $ T=0 $.  
We treat at first the Green's function for time ordering
 $\tau -\tau '>0$. Then the path integral (\ref{300}) can be interpreted
 as follows: \\
 With the help of
 \begin{eqnarray}
  \lefteqn { H_{ss}(\vec{a}_{CS},\vec{A}^t_{ss,2}) }\label{340}\\   
   & = &    \int d^2r \Bigg\{
 \frac{1}{2m}\Psi^+(\vec{r}) \bigg(-i\vec{\nabla}+\vec{A}  
 + \vec{a}_{CS}+\vec{A}^t_{ss,2}\bigg)^2\Psi(\vec{r}) \nonumber \\
 & & + \frac{1}{2} \nts{1} \int d^2r' 
 (|\Psi(\vec{r})|^2-\rho_B)
 V^{ee}_{\vec{r},\vec{r}'}(|\Psi(\vec{r}\,')|^2-\rho_B) \Bigg\}  \nonumber  
 \end{eqnarray}
and
\begin{equation}
 \vec{A}^t_{ss,2}=  \left(\frac{1}{2}-\frac{t}{\beta}\right)
 \tilde{\phi} \vec{f}(\vec{r}-\vec{x})- \left( \frac{3}{2}-
   \frac{t}{\beta}\right)
 \tilde{\phi} \vec{f}(\vec{r}-\vec{x}\,')\label{350}
 \end{equation}
 we define the time evolution operator
 $ U_t(\vec{a}_{CS},\vec{A}_{ss,2}) $ by
\begin{equation} 
-\frac{\partial}{\partial t}U_t(\vec{a}_{CS},\vec{A}_{ss,2})=
H_{ss}(\vec{a}_{CS},\vec{A}^t_{ss,2})\,
U_t(\vec{a}_{CS},\vec{A}_{ss,2})\label{360}
\end{equation}
with  the boundary condition $ U_0 = 1$.
Because the flux quantum numbers of the strings of the vector potentials 
$ \vec{A}^\beta_{ss,2} $ and $\vec{A}^0_{ss,2}  $
differ only by an integer value we know that the eigenspaces of
$ H_{ss}(\vec{a}_{CS},\vec{A}^0_{ss,2}) $ and
$ H_{ss}(\vec{a}_{CS},\vec{A}^\beta_{ss,2}) $ are
linked by the unitary phase transformation  
\begin{eqnarray}
 \mbox{Ph}_{\tilde{\phi}}(\vec{x},\vec{x}\,') & = & \exp[-i\tilde{\phi} \int d^2r
\, \mbox{\small arg}(\vec{x}-\vec{r}) \hat{\rho}(\vec{r})]  \label{370}  \\
& &  \times \exp[i\tilde{\phi} \int d^2r
\, \mbox{\small arg}(\vec{x}\,'-\vec{r}) \hat{\rho}(\vec{r})]       \nonumber 
\end{eqnarray}
acting on the states of $ H_{ss}(\vec{a}_{CS},\vec{A}^\beta_{ss,2}) $. 
The eigenvalues are invariant
 under this transformation. 
With the help of the phase factors (\ref{370}) we get
 for the Green's function for $ \tau-\tau'>0$  
  \begin{align}
& G^{\mbox{\scriptsize{Weyl}}}(\vec{x},\vec{x}\,';\tau,\tau')=
e^{-i \tilde{\phi} \vec{f}(0)(\vec{x}-\vec{x}\,')}
\label{380}  \\
&   
\times \lim\limits_{\beta \to \infty}
  \frac{1}{\mbox{Tr}\left[\exp\left[-(\beta+\tau-\tau') H_{ss}(\vec{a}_{CS},0)\right]\right]} \nonumber  \\
& \times \mbox{Tr}\Big[U_\beta(\vec{a}_{CS},\vec{A}_{ss,2})
   \mbox{Ph}_{\tilde{\phi}}(\vec{x},\vec{x}\,')
    \nonumber  \\
& \times \Psi_{e}(\vec{x})\exp[-(\tau-\tau')
H_{ss}(\vec{a}_{CS},\vec{A}^\beta_{ss,2})]\Psi^+_{e}(\vec{x}\,')
\Big] \nonumber 
 \end{align}
where $ \Psi^+_e $ is the electron creation operator (\ref{5}).  
Now assume that $ H_{ss}(\vec{a}_{CS},\vec{A}^t_{ss,2}) $
has the same ground state degeneracy and ground state energy as the
Hamiltonian 
$ H_{\mbox{\scriptsize ss}}(\vec{a}_{CS},0) $.
This will be shown for the non-interacting as well as for the
interacting electron system in the following subsections. 
Then it is intuitively clear that the operator   
$ U_\beta(\vec{a}_{CS},\vec{A}_{ss,2}) /
\mbox{Tr}[\exp [-(\beta+\tau-\tau') H_{ss}(\vec{a}_{CS},0)]] $
has its support on the lowest energy eigenspace  for $ \beta \to
\infty $. We will also show this explicitely in the following subsection.
On the other hand it is clear that we get an infinite
or zero Green's function, respectively, if the integral of
the ground state energy of $ H_{ss}(\vec{a}_{CS},\vec{A}^t_{ss,2})-
 H_{\mbox{\scriptsize s}}(\vec{a}_{CS},0)  $ over the time $ t $ is non-zero.
As mentioned above  the incorrect reproduction of the ground state energy
within second order perturbation
theory is the reason that the RPA self energy is infinite for $ T=0 $. 

Now we compare the expression (\ref{380}) of the Green's function 
with the
corresponding expression in the Coulomb
gauge for $ \tau-\tau'>0$. This function is given by the expression
(\ref{380}) with the substitutions
$ U_\beta(\vec{a}_{CS},\vec{A}_{ss,2}) \mbox{Ph}(\vec{x},\vec{x}\,')
\rightarrow U_\beta(\vec{a}_{CS},0) $, $  \Psi_{e}(\vec{x}) \rightarrow
\Psi( \vec{x}) $, $  \Psi^+_{e}(\vec{x}) \rightarrow \Psi^+(  \vec{x}) $,
$ H_{ss}(\vec{a}_{CS},A_{ss,2}^\beta) \rightarrow
H_{ss}(\vec{a}_{CS},0)$ and $ \exp[-i \tilde{\phi}
\vec{f}(0)(\vec{x}-\vec{x}\,')]  \rightarrow 1$. As mentioned above 
we calculated  in \cite{di3} a CS Green's function in the Coulomb gauge that
vanishes exponentially with an exponent proportional to $ \log(A) $.
The reason for this is that a quasi particle $ \Psi^+(  \vec{x}\,')$
created at time $ \tau' $ 
gets a velocity boost from all other particles in the ground
state. This results in an infinite quasi particle energy. This is not    
the case for the quasi particle $ \Psi^+_{e}(  \vec{x}\,')$
created at time $ \tau' $ 
in the Green's function of the temporal gauge.
Thus, we see that the  CS phases between the created particle
and all other electrons
of the ground state which led 
to the velocity boost in the Coulomb gauge are automatically
annihilated through the dynamical creation of an opposite phase by turning on the magnetic
strings in $ H_{\mbox{\scriptsize ss}}$ in the
temporal gauge.
In this sense one can understand the missing $ \log(A) $
terms in the RPA self energy of the temporal gauge. Now we carry out a CS
retransformation (\ref{5}) of the expression (\ref{380}). This results in
  \begin{align}
& G^{\mbox{\scriptsize{Weyl}}}(\vec{x},\vec{x}\,';\tau,\tau')=
e^{-i \tilde{\phi} \vec{f}(0)(\vec{x}-\vec{x}\,')}
\label{390}  \\
&   
\times \lim\limits_{\beta \to \infty}
  \frac{1}{\mbox{Tr}\left[\exp\left[-(\beta+\tau-\tau') H_{ss}(0,0)\right]\right]} \nonumber  \\
& \times \mbox{Tr}\Big[U_\beta(0,\vec{A}_{ss,2})
   \mbox{Ph}_{\tilde{\phi}}(\vec{x},\vec{x}\,')
    \nonumber  \\
& \times \Psi(\vec{x})\exp[-(\tau-\tau')
H_{\mbox{\scriptsize ss}}(0,\vec{A}^\beta_{ss,2})]\Psi^+(\vec{x}\,')
\Big]\,. \nonumber 
 \end{align}
In the following subsection we will calculate this expression for the case of
a non-interacting electron system. 
Then we see from (\ref{390}) that by the CS retransformation one
loses all many-particle interaction terms. Thus, we can determine the
Green's function by calculating the one particle 
ground state wave function and the ground state energy of the Hamiltonian
$ H_{ss}(0,\vec{A}^t_{ss,2}) $. From this we get   
$ U_\beta(0,\vec{A}_{ss,2}) $. In subsection A, 
we calculate the ground state wave function and the ground
state energy of an electron in a homogenous magnetic field in the background of
two opposite magnetic strings. By using this result,
we calculate
in subsection B the Green's function in the temporal gauge without Coulomb
interaction. Subsection C is devoted to show that the Green's function
of the interacting electron system should be  also finite.

\subsection{The quantum 
  mechanics of an electron in a homogeneous magnetic field
  and a background of one or two seperated magnetic strings}
In this subsection we discuss the eigenfunctions and eigenvalues of an electron
in a homogenous magnetic field with a background of one or two seperated
strings. The eigenvalues and eigenfunctions of these systems have to be
calculated to get an expression for the one particle propagator
$ U_\beta(0,\vec{A}_{ss,2}) $  (\ref{360}) 
as well as for the exact density-density propagator $ {\cal
  D}^{\mbox{\scriptsize ex}}_{00} $ 
(\ref{280}). \\

We will
solve at first the simpler problem of an electron
in a homogeneous magnetic field
and a background of {\it one} magnetic string at the origin with flux quantum
$ \phi$. The one-particle Hamiltonian is given by
\begin{equation}
H^1_s(\phi)= \frac{1}{2m}  \left(-i \vec{\nabla}+\vec{A}(\vec{r}) -
  \phi\vec{f}(\vec{r}) \right)^2 \;. \label{392}
\end{equation}
We now seek a solution of the  the form
$ \Psi(r,\varphi)=\frac{1}{\sqrt{2 \pi}} f(r) e^{i p \varphi} $ (we used 
cylindrical polar coordinates). Then one gets for the eigenvalue equation
in polar coordinates with $ \vec{A}=B (y,-x)/2 $ 
\begin{eqnarray}
& \nts{40} \frac{1}{2m}\left(f''+\frac{1}{r}f' -\frac {(p -
   \phi)^2}{r^2} f \right)  \label{394} \\
&  +\left(E- \frac{1}{8} m \omega_c^2 r^2 +\frac{1}{2} \omega_c (p -
    \phi)\right) f=0 \,. \nonumber 
\end{eqnarray}  
This differential equation is similar to the differential equation of
electrons in a homogeneous magnetic field $ B $ without a magnetic string
which can be recovered by replacing $ p - \phi \rightarrow p $. The 
eigenfunctions  and eigenvalues of this system are well known 
(e.g.\cite{cha1}). By using an analogous method to solve the  differential
equation (\ref{394}) we get for the regular eigenfunctions  and
eigenvalues which are finite at $ r=0 $ 
 \begin{eqnarray}
   \Psi^\phi_{n, p}(\vec{r}) & = &  \left[\frac{n !}{2 \pi l_0^2 2^{|p - \phi|}
   \Gamma(n+1+|p - \phi|)}\right]^{\frac{1}{2}}
 e^{i p \phi } \,,  \label{396}   \\
& & \times \left(\frac{r}{l_0}\right)^{|p - \phi|} 
 L_n^{|p - \phi|}\left(\frac{r^2}{2 l_0^2}\right) e^{-\frac{r^2}{4 l_0^2}}
 \nonumber \\
 E^\phi_{n, p} & = & \omega_c \left(n + \frac{1}{2}|p - \phi|-
   \frac{1}{2}(p - \phi)+ \frac{1}{2}\right)   \label{398}
 \end{eqnarray}
 with $ n \in \mathbb{N}_0 $ and $ p \in \mathbb{Z} $.  
Beside these eigenfunctions we also have eigenfunctions which are not
finite at $ r=0 $ (but nevertheless square integrable).
These irregular eigenfunctions and eigenvalues are given by
(for $ 0 \le \phi \le 1$)  
 $ \Psi^{\phi,\mbox{\scriptsize sing}}_{n, 0 }  \propto 
 (r/l_0)^{-\phi}
 L_n^{-\phi}(r^2/(2 l_0^2)) e^{-r^2/(4 l_0^2)}$ with eigenvalues
 $ E^{\phi,\mbox{\scriptsize sing }}_{n, 0}=\omega_c(n+1/2)$
 and $ \Psi^{\phi,\mbox{\scriptsize sing}}_{n,1}
     \propto e^{i \varphi} 
 (r/l_0)^{-(1-\phi)}
 L_n^{-(1-\phi)}(r^2/(2 l_0^2)) e^{-r^2/(4 l_0^2)}$ with eigenvalues  
$ E^{\phi,\mbox{\scriptsize sing}}_{n, 1} =\omega_c
 (n+\phi-1/2) $.
It is well known for an electron in the background
of a magnetic string \cite{bo1,aud1}  that there is no domain of the
Hamiltonian $ H^1_s $ which contains the eigenfunctions 
$ \Psi^\phi_{n, p}(\vec{r})  $ as well as the singular eigenfunctions 
$ \Psi^{\phi,\mbox{\scriptsize sing}}_{n, 0 } $ and
$ \Psi^{\phi,\mbox{\scriptsize sing}}_{n, 1}$
in a way that the operator $ H^1_s $ is self adjoint. This is the reason
for the non-orthogonality of the regular and the irregular
eigenfunctions \cite{aud1}. It is now possible to restrict the domain of the
Hamiltonian to get a self adjoint extension of $ H^1_s $. This restriction
is not unique. The concrete  extension has to be determined by physical
arguments. For example in the Aharonov-Bohm case \cite{ah1} (i.e. electrons
in the background of a magnetic string) one can show rather 
generally that the correct self adjoint
extension of the Hamiltonian consists of the domain of wave functions which
are zero at the origin \cite{hag1} (for $ 0<\phi<1$). This is done
by a regularization of the
magnetic string field at the origin (the string width $ R_0 $ being
finite). After a calculation of the inner and outer solutions of this spread
out string and a calculation of the matching conditions one gets for $ R_0 \to
0 $ only a square integrable
non-zero eigenfunction in the case where the function is
zero at the origin (for $ 0<\phi<1$). We have done a similar 
calculation for the case of an
electron in the background of a 
finite number of homogeneously spread out magnetic strings 
in a homogeneous magnetic field.
It is then easily seen that the asymptotics of the wave function at the border
$ R_0 $ of a string does not depend on the existence of the homogeneous
magnetic field and the other strings for $ R_0 \to 0 $
(this could be also seen from equation (\ref{394})). By an examination of the
matching conditions we get a square integrable non-zero 
eigenfunction only in the case
where the function is zero at the origin of the strings.
Thus, we have to use the regular eigenfunctions (\ref{396}) as solutions.
One can see from these eigenvalues that in
contrast to the case of an electron in
a string background without a homogeneous magnetic field we have an
energy splitting of the Landau levels due to the string background. 

Next, we will calculate the eigenfunctions and eigenvalues of an
electron in a homogeneous magnetic field $ B $ in the background of 
{ \it two }
magnetic strings of opposite strength separated by a distance
$ d $. In contrast
to the one string system above we do not have a rotational
symmetry. This makes it much more complicated to get the eigenfunctions and
eigenvalues of the system. Therefore, we will restrict
us in the following to the ground state.
This is enough because one can calculate
the Green's function for $ T=0$ from the
knowledge of the ground state
eigenfunctions and eigenvalues 
due to the denominator in (\ref{390}). The Hamiltonian of an
electron in a homogeneous magnetic field in the background of
two magnetic strings
with flux quanta  $ -\phi $ and $ \phi $ located
at the origin and $ d \vec{e}_x $ is given by
\begin{equation}
H^1_{ss}(\phi)= \frac{1}{2m}  \left(-i \vec{\nabla}+\vec{A}(\vec{r}) -
  \phi\vec{f}(\vec{r})+
  \phi\vec{f}(\vec{r}-d\vec{e}_x ) \right)^2\,. \label{400}
\end{equation}
We suppose that $ \phi \ge 0 $. 
We now carry out a phase transformation on the eigenfunctions $ \Psi $
of $ H^1_{ss} $. It is  given by
\begin{equation}
  \Psi_p(\vec{r})=e^{-i \phi \arg[\vec{r}]}
  e^{+i \phi \arg[\vec{r}-d\vec{e}_x]}\Psi(\vec{r})\,. \label{410}
\end{equation}   
In the following we denote the one dimensional subspace
$ y=0 $, $ 0 \le x \le d \vec{e}_x $ of the plane by $ C $.
With the help of the transformation (\ref{410}) one easily can show that
$ \Psi_p $ are eigenfunctions of the Hamiltonian without the strings 
$ H^1 = (-i \vec{\nabla}+\vec{A}(\vec{r}))^2/2m $ for $ r=\mathbb{R}^2 \backslash C $
with the matching conditions 
\begin{eqnarray}
   \Psi_p(\vec{r}+\epsilon \vec{e}_y) & = & e^{i \phi 2 \pi}
   \Psi_p(\vec{r}-\epsilon \vec{e}_y)\, , \nonumber \\
 \vec{\nabla} \Psi_p(\vec{r}+\epsilon \vec{e}_y) & = & e^{i \phi 2 \pi}
  \vec{\nabla} \Psi_p(\vec{r}-\epsilon \vec{e}_y)   \label{420}
\end{eqnarray}
for $ \vec{r} \in C $ and $ \epsilon \to 0^+ $.
By using the
complex variables $ z=x+iy $ and $ \overline{z}=x-iy $ we get for  
the Hamiltonian $ H^1 $
\begin{equation}
H^1=-\frac{2}{m} \partial_z \partial_{\overline{z}}- \frac{B}{2m}\left(
    z  \partial_z-\overline{z}  \partial_{\overline{z}}\right) +
  \frac{B^2}{8m} \overline{z}z \,.  \label{430}
\end{equation}  
With the help of the ansatz $ \Psi_p(z, \overline{z})=
u(z, \overline{z})e^{-|z|^2 B/4} $ we have to solve
on $ \mathbb{R}^2 \backslash C $ the eigenvalue
equation $ H^1_z u(z,\overline{z}) =E
u(z,\overline{z}) $ with
\begin{equation}
H^1_z=-\frac{2}{m} \partial_z \partial_{\overline{z}}+ \frac{B}{m}
\left(\overline{z}  \partial_{\overline{z}}\right) +
  \frac{B}{2m}   \label{440}
\end{equation}  
and the scalar product
\begin{equation}
\langle u,v \rangle_1=\int d^2r \;e^{-|z|^2 B/2} \;\overline{u} v \label{445} 
\end{equation}
for two wave functions $ u, v $. 
The energy of a normalized wave function $ u $ on $ \mathbb{R}^2\backslash C
 $ is given by
\begin{eqnarray}
E & = & \langle u(z, \overline{z}), H^1_z(z, \overline{z})
u(z, \overline{z})\rangle_1  \label{450}  \\
 & =   & \frac{2}{m} \langle \partial_{\overline{z}}u(z, \overline{z}),
 \partial_{\overline{z}}u(z, \overline{z}) \rangle_1 +\frac{B}{2m}
 \nonumber 
\end{eqnarray}
From this equation we obtain
that the ground state wave functions are the normalized holomorphic wave
function on $ \mathbb{C}\backslash C $ which fulfill the transformed matching
conditions corresponding to (\ref{420}) (here we identified
$ \mathbb{R}^2 $ with the complex plane $ \mathbb{C} $).

We now determine a linearly independent basis
of the ground state wave functions. To this end we carry  out the
following transformations on the ground state wave functions
\begin{equation}
  u_t(z)=z^{\phi}(z-z_0)^{-\phi}
  u(z)\,.  \label{460}
\end{equation}
Here $ z_0=d $ and $ z \in \mathbb{C}$.
The scalar product of the transformed wave functions is given by
\begin{equation}
\langle u_t,v_t \rangle_2=\int d^2r \; e^{-|z|^2 B/2}
|z|^{-2 \phi}|z-z_0|^{2\phi}
\;\overline{u}_t v_t \,.  \label{470} 
\end{equation}
Using the matching conditions (\ref{420})
we obtain that the ground state wave functions are the
holomorphic functions on the whole complex plane $\mathbb{C}  $
with a finite norm 
corresponding to the scalar product (\ref{470}).
It is shown in \cite{ha1} that this space as well as the space of the
holomorphic functions on $ \mathbb{C} $ with a finite norm (\ref{445})
known as the Segal-Bargmann space \cite{ba1} are Hilbert spaces.
We will denote the first by $ HL^2_2 $ and the Segal-Bargmann space
by $ HL^2_1 $.
One can see easily that both spaces consists of the same holomorphic
functions. It is well known that the functions $ \{z^p\} $ ($ p \in
\mathbb{N}_0 $)
are a basis  \cite{ha1} of the Segal-Bargmann space $ HL^2_1 $.
We obtain from the definition  
of the scalar product (\ref{445}) that the basis functions
are orthogonal. This is no longer the case for $ \{z^p\} $
($ p \in \mathbb{N}_0 $)
in the Hilbert space with the scalar product (\ref{470}). Nevertheless,
we will show in Appendix C that these functions are indeed a basis of this
Hilbert space.  

Then, by carrying out the retransformations (\ref{460}) and (\ref{410})
we get a basis for the ground state eigenfunctions  of  $ H_{ss}^1 $ (\ref{400}) for $ 0 \le
\phi \le 1$. Due to the considerations below (\ref{398}) concerning the
domain of the  Hamiltonian we find for this basis  
\begin{equation}
\Psi_p^{\phi}  =  e^{i p \varphi} r^{p-\phi} 
\sqrt{r^2-2d r \cos{\varphi}+d^2}^{\phi}e^{-r^2 \frac{B}{4}}
\; \mbox{for} \; p \ge 1,  \label{480} 
\end{equation}
where $ 0 \le \phi \le 1$.
We find one additional basis state for the limiting cases $ \phi=0,1 $ 
\begin{equation}
 \Psi_0^{\phi}= e^{-r^2 \frac{B}{4}} \qquad  \mbox{for} \;
 \phi=0,1 
 \label{483} 
\end{equation}   
We see from (\ref{450}) that the energy eigenvalues $ E_{p} $ 
of the wave functions $ \Psi_p^{\phi} $ (the ground state energy for 
string strength $ \phi $) are given by
\begin{equation} 
E_{p}=\frac{\omega_c}{2} \,.\label{485} 
\end{equation}
By comparing the ground state degeneracy as a function of
$\phi$ and the energy of an electron in a homogeneous magnetic field in the
background of one string (\ref{396}), (\ref{398}) and in the background of
two magnetic strings (\ref{480}), (\ref{483}), (\ref{485}) we get agreement of these two
systems for string distance $ d \to \infty $. 
In Fig. 1 we show the ground state energy as a function of the
magnetic flux for the Hamiltonian of an electron in the background of two
magnetic strings. We see from this figure that only
one wave function ($ p=0 $) 
of the degenerate ground state for $ \phi=0 $ increases in energy.
The rest of the ground states ($ p \not =0 $) keep their lowest Landau
level energy.

\begin{figure}[t]
  \psset{unit=3cm}
\begin{pspicture}(2.3,1)
 \psset{linewidth=0.5pt,arrowinset=0,linecolor=black} 
\psaxes[Dy=2,dy=2,Dx=1,dx=2,Oy=0.0,labels=x]{->}(2.2,1.0)
\psline[linewidth=1.7pt,linecolor=gray](0.0,0.25)(2.0,0.25)
\psline[linewidth=3 pt,linecolor=black](-0.05,0.25)(0.05,0.25)
\psline[linewidth=3 pt,linecolor=black](-0.05,0.75)(0.05,0.75)
\psline[linewidth=3 pt,linecolor=black](1.95,0.25)(2.05,0.25)
\psline[linewidth=3 pt,linecolor=black](1.95,0.75)(2.05,0.75)
\psline[linewidth=0.3pt,linecolor=gray](0.0,0.25)(2.0,0.75)
\psline[linewidth=0.3pt,linecolor=gray](0.0,0.75)(2.0,0.25)
\rput(-0.15,0.25){$ 0.5 $}
\rput(-0.15,0.75){$ 1.5 $}
\rput(-0.18,1.02){$ E [\omega_c]$}
\rput(2.15,-0.1){$ \phi$}
\rput(0.15,0.17){$N_{LLL}  $}
\rput(2.00,0.17){$N_{LLL}$}
\rput(1.0,0.17){$N_{LLL}-1$}
\end{pspicture}
\vspace*{0.3cm}
\caption{Ground state energy $ E $ of an electron in a magnetic field
  $ B $ in the background of two magnetic strings with flux $ \phi $
  and $- \phi $. In the figure we show $ E $ as a function of the flux
  $\phi$. $ N_{LLL}$ is the number of lowest Landau level eigen states 
for $\phi=0$. This number is proportional to the area of the system. 
\label{fig1}} 
\end{figure}
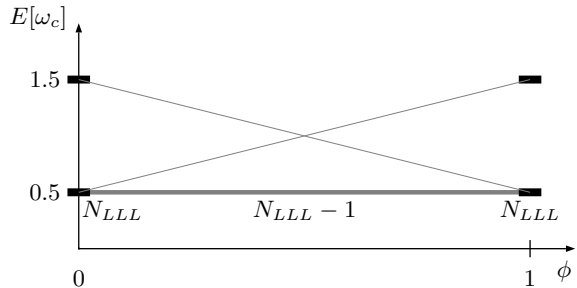

\subsection{The calculation of the Green's function}
In this subsection, we will calculate the Green's function
$ G^{\mbox{\scriptsize Weyl}}(\vec{x},\vec{x}\,';\tau,\tau')  $ (\ref{380}) for
$ \tau-\tau'>0 $ and $ \beta \to \infty $.
Afterwards we will generalize
the results to times $ \tau-\tau'<0 $. For doing this
  we have  to calculate the thermodynamic time evolution $ U_\beta $ of the
 time dependent Hamiltonian $ H_{ss}(0,\vec{A}^t_{ss,2})$ (\ref{340}).
 We first will calculate the one-particle transition matrix
 of the adiabatic time dependent Hamiltonian
 $ H^1_{ss}((t/\beta) \tilde{\phi}) $ ($ t $ is the time parameter)
 divided by 
 the one-particle partition function $ \mbox{Tr}[ \exp[-\beta
 (H^1_{ss}(0)-\mu)]] $
 for $ \beta \to \infty $.
 We will denote this quantity by $ \mbox{UP}^1(\phi) $ where
 $ \phi=(t/\beta) \tilde{\phi} $. 
  This can be calculated from the finite wave functions
  for $ \beta \to \infty $
 of the transformed Schr\"odinger wave equation
 \begin{equation}
  \partial_{\phi} \varphi(\phi) =-
  \frac{\beta}{\tilde{\phi}} \, \left(H^1_{ss}(\phi)-\frac{\omega_c}{2}\right)
    \varphi(\phi)    \label{490} 
\end{equation}
for $ 0 \le \phi \le \tilde{\phi} $.
Then the solutions of the Schr\"odinger
equation of motion for imaginary times of the Hamiltonian
$ H^1_{ss} $ are given by 
 $ \Psi(\phi) = \exp[-(\beta/\tilde{\phi})
 (\omega_c/2)\phi]\varphi(\phi)$.  We will solve
 equation (\ref{490}) first
for $ 0 \le \phi \le 1 $ and $ \beta \to \infty $.  
We now define $ \varphi=\varphi_0+\varphi_A $ where 
$ \varphi_0 $ is that part of the wave function $ \varphi $
which is in the
eigenspace $ H^1_{ss} $ of the lowest eigenvalue $ \omega_c/2 $ for $ \phi
\not=0 $ and $ \phi \not=1 $
(i.e. $\mbox{span}\{\Psi_p^{\phi}\} $ for $ p \ge 1$).
For $ \phi=0 $ or
$ \phi=1 $, respectively, we denote $ \varphi_0 $ by that part of
the ground state wave function which is linked continuously as a function 
of $ \phi $ to a ground state wave function
for $ 0 < \phi  $ or $ \phi <  1 $,
respectively. Thus, $ \varphi_A $ is that part of the wave function
$ \varphi $ which is in the subspace spanned by the higher energy
eigenfunctions. 

In appendix C we show that 
$ \langle \varphi_A| \varphi_A \rangle  (\phi) $ vanishes as
$ O( 1/\beta ) $ for $ 0 <\phi<1 $ and as
$ O( 1/\beta^{1/(n+1)} ) $ at $ \phi=1 $ where $ n $ is defined by the
order of the intersection of the eigenvalues at $ \phi=1 $.
In the case of a smooth intersection we find  that 
$ \langle \varphi_A| \varphi_A \rangle  (\phi) $ vanishes as
$ o(1) $ at $ \phi=1 $.
This corresponds  to the well known adiabatic theorem \cite{ka1}
in the case of the solutions of the Schr\"odinger equation (\ref{490})  for
real times. 
Furthermore, we obtain in appendix C  that the transition operator
 $ \mbox{UP}^1(\phi) $ is non-zero only 
 on the subspace spanned by $ \Psi^0_p $ (\ref{480}) for $ p \ge 1 $.
 One can then get the finite part of $ \mbox{UP}^1(\phi) $
 by solving the projected Schr\"odinger equation
\begin{equation}
\langle \Psi^{\phi}_p|\partial_{\phi}
\varphi_0(\phi)\rangle=0 \label{500}
\end{equation} 
for $ p=1 \dots  \infty $.  The set of equations (\ref{500}) shows that
the time evolution of $\varphi_0(\phi) $
is given by a parallel  transport in the sub manifold of the ground states.
This is a well known transport in quantum mechanics 
which is responsible for the Berry phase
\cite{be1} in the case of a non-degenerate ground state.
Supposing that $ \varphi_0(0)=\Psi^{0}_p $, we get
$ \mbox{UP}^1(1)\varphi_0(0)=\varphi_0(1) $
for the evolution where the background strings are switched on adiabatically
up to one flux quantum.
In the following, we will calculate this quantity for small string separation
$ d $. 
At first, we have to calculate the overlap matrix $ S_{nl}^d$:
\begin{eqnarray}
& & S^d_{nl}(\phi):=\int d^2r \Psi^{\phi}_n(\vec{r})
\Psi^{\phi}_l(\vec{r})  \label{510}  \\
& & = \frac{2 \pi 2^n n!}{B^{n+1}}\Bigg[\delta_{nl}\left(1+
    \frac{\phi^2 B}{2n}d^2
  \right) -\delta_{n+1,l} \phi d -\delta_{n-1,l}\frac{\phi B}{2n}d  \nonumber \\
& & 
    + \delta_{n+2,l}O(1/n^0)O(d^2)+\delta_{n-2,l}O(1/n^2)O(d^2)+O(d^3) \Big] \nonumber 
\end{eqnarray}  
With the help of this overlap matrix it is easy to solve
the parallel transport equation (\ref{500}). Defining
the matrix $ [\mbox{UP}^1](\phi)$ by
$ \vec{c}(\phi)  = [\mbox{UP}^1](\phi) \vec{c}(0) $ where
$ \varphi_0(\phi)=\sum_p c_p(\phi) \Psi_p^{\phi} $ 
we get as a solution of (\ref{500}) 
\begin{align}
& [\mbox{UP}^1](\phi)= 
T \exp\left[-\frac{1}{2} \int\limits_0^{\phi}
   d\phi' (S^d(\phi'))^{-1}(\partial_{\phi'} S^d(\phi'))
 \right]    \nonumber  \\
 &= T \exp\Bigg[-\frac{1}{2} \int\limits_0^{\phi}
   d\phi' \left(1+(S^0(\phi'))^{-1}\Delta S^d(\phi')\right)^{-1} \nonumber \\
&  \qquad \qquad  \qquad\times (S^0(\phi'))^{-1}\partial_{\phi'}\Delta
S^d(\phi') \Bigg]\,.  \label{520}
\end{align}
Here, $ S=S^0+\Delta S^d $ where $ S^0 $ is the overlap matrix $ S^d $
for $ d=0 $. 
Now we expand the exponential function  in (\ref{520}) and the term
$ (1+(S^0)^{-1}\Delta S^d)^{-1} $ in the exponent. Then  we get an expansion
of $ \mbox{UP}^1(1) $ in $ d $. 
$ \mbox{UP}^1(\tilde{\phi})$ is then given by  
$ \mbox{UP}^1(\tilde{\phi})=(\mbox{Ph}^{1}_{-\tilde{\phi}}(0,d \vec{e}_x))
(\mbox{Ph}^1_1(0,d \vec{e}_x)
\mbox{UP}^1(1))^{\tilde{\phi}} $. 
$ \mbox{Ph}^1 $ is the phase transformation operator (\ref{370}) 
calculated in the one-particle sector in the first quantized language.
By using $ \mbox{UP}^1(1) \Psi_0^{0}=0 $ we get
\begin{align}
& [ \mbox{Ph}^1_{\tilde{\phi}}(0,d \vec{e}_x)\mbox{UP}^1(\tilde{\phi})]_{nl}
=(1-\delta_{n0})(1-\delta_{l0}) \label{540}  \\
&  \times \Bigg\{\delta_{nl}\left(1-\frac{\tilde{\phi}^2 B}{16}
 \frac{(2n+1)}{n(n+1)} d^2
\right)+\delta_{n1} \delta_{l1} \frac{\tilde{\phi}( \tilde{\phi}-1)B}{16n}
d^2        \nonumber \\
&+ \left(\delta_{n-1,l} \frac{\tilde{\phi} B}{4 n } \sqrt{2n B} -\delta_{n+1,l} \frac{\tilde{\phi}}{2}
  \frac{1}{\sqrt{2(n+1) B}} \right)d 
\Bigg\} \nonumber \\
& -\delta_{n,0}\, \delta_{l,1} \, \frac{1}{\sqrt{2B}}\, d \nonumber \\
&  + \delta_{n+2,l}O(1/n^1)O(d^2)+\delta_{n-2,l}O(1/n^1)O(d^2)+O(d^3)
    \nonumber  
\end{align}
where $ [\mbox{Ph}^1_{\tilde{\phi}}
(0,d \vec{e}_x)\mbox{UP}^1(\tilde{\phi})]_{nl} $ is the matrix
operator with respect to
the orthonormal basis $ \{ \Psi_n^{0} \} $ (\ref{480})
for  $n=0 \ldots \infty$.  
With the help of this matrix we will calculate the Green's
function $ G^{\mbox{\scriptsize Weyl}} $
(\ref{390}).  
For doing this we first calculate the overlap function
$  \mbox{Ov}(d) $ by the use of this matrix yielding
\begin{align}
 &   \mbox{Ov}(d) :=
 \frac{\sum\limits_{ \Psi_{\vec{p}}\in \nu_{1/\tilde{\phi}},p_i \not= 0} 
 \mbox{Det}\left[\mbox{Ph}_{\tilde{\phi}}^1(0,d \vec{e}_x)\mbox{UP}^1(\tilde{\phi})\right]_{\vec{p}}}{
 \sum\limits_{ \Psi_{\vec{p}}\in \nu_{1/\tilde{\phi}}} 1} \nonumber \\
&   =  \frac{\sum\limits_{ \Psi_{\vec{p}}\in \nu_{1/\tilde{\phi}},p_i \not= 0} 
 \exp\left[\mbox{Tr}\left[\log\left[
       \mbox{Ph}^1_{\tilde{\phi}}(0,d \vec{e}_x)
       \mbox{UP}^1(\tilde{\phi})
  \right]_{\vec{p}}\right]\right]}{
 \sum\limits_{ \Psi_{\vec{p}}\in \nu_{1/\tilde{\phi}}} 1}
\nonumber \\
&  = \nts{2} \left( \nts{1} 1-\frac{1}{\tilde{\phi}}
\nts{1} \right) \nts{1} 
\exp\left[-(\tilde{\phi}-1) \, \frac{B}{8} \,\log(A) \, d^2 +O(1/A^0)O(d^2)
  \right] \nts{2} .
\label{550}
\end{align}
Here $ \nu_{1/\tilde{\phi}} $ are
the ground states of the $ \nu=1/\tilde{\phi} $ system.
The ground state $ \Psi_{\vec{p}} $ is then given by the Slater determinant
 $ \Psi_{\vec{p}}=\mbox{S}[\Psi^{0}_{0,p_i},\ldots,\Psi^{0}_{0,p_N}] $.
The wave functions $ \Psi^{0}_{0,p_N} $ can be seen in equation (\ref{396}).   
$ [\cdot]_{\vec{p}} $ is the sub matrix  of the argument 
with line and column indices $ p_i $.
For deriving this result we used techniques of averaging over the
ground states developed in \cite{di2}.
In (\ref{550}) we mean by $ O(1/A^0)O(d^2) $ that these terms are finite for
$ A \to \infty  $ and of order $ O(d^2) $.
The finiteness of the exponents of order  $ O(d^3) $ for $ A \to \infty $
is obtained by extending the analysis of the overlap matrix
$ S_{nl}^d $ (\ref{510}) as a function of $ n,l $ to all orders of $ d $.
We see from (\ref{550}) that
the overlap function $ \mbox{Ov}(d) $ is zero for $ d > 0 $ and 
$ A \to \infty $. This is in accordance with the general inequality
$ -1 \le  \mbox{Ov}(d) \le 1 $ that  can be derived from the
parallel transport equation (\ref{500}) by showing that the norm of
the vector $ \varphi_0 $ is invariant under the transport. 

It is now easy to calculate the Green's function
$ G^{\mbox{\scriptsize Weyl}}(\vec{x}, \vec{x}\,'; \tau, \tau') $
(\ref{390}) for $ \tau-\tau'>0$.
This is done by writing down the expression (\ref{390})
in the one-particle basis.
With the help of (\ref{540}), (\ref{550}) and by using 
the above mentioned averaging techniques
as well as (gauge) transformation properties of the path integral
(\ref{140}) to get the Green's function also for $ \vec{x} \not= 0 $ we obtain  
\begin{align}
& G^{\mbox{\scriptsize Weyl}}(\vec{x}, \vec{x}\,'; \tau, \tau')=
 (-1)^{\frac{\tilde{\phi}}{2}}
 \mbox{Ov}(|\vec{x}-\vec{x}\,'|)\,e^{-i \tilde{\phi}
 \vec{f}(0)(\vec{x}-\vec{x}\,')} \nonumber \\ 
& \times \Bigg(
G^0(\vec{x}, \vec{x}\,'; \tau, \tau')
+ e^{-i\vec{A}(\vec{x})(\vec{x}-\vec{x}\,')}
e^{-|\vec{x}-\vec{x}\,'|^2 \frac{B}{4}}
\label{560} \\
& \times 
e^{-(\tau-\tau') \left(\frac{\omega_c}{2}-\mu\right)}
    \left(\frac{1}{\tilde{\phi}}\frac{B}{4 \pi}\, |\vec{x}-\vec{x}\,'|^2
      +O(B^2|\vec{x}-\vec{x}\,'|^4)\right)\Bigg). \nonumber 
\end{align}
$ G^0 $ is the Green's function of the Hamiltonian $ H_{ss}(0,0) $,
calculated with respect to the vacuum ground state (zero particle ground
state), i.e. $ G^0(\vec{x}, \vec{x}\,'; \tau, \tau')=
\Theta(\tau-\tau')\int d^2r \, \delta(\vec{x}-\vec{r})
\exp[-(\tau-\tau') ( 1/(2m) (-i \vec{\nabla}+\vec{A}(\vec{r}))^2-\mu) ] \,
\delta(\vec{x}\,'-
\vec{r})$.   
One can deduce from (\ref{300}) that the expression (\ref{560}) 
is also valid for the Green's function $ G^{\mbox{\scriptsize Weyl}}(\vec{x},
\vec{x}\,'; \tau, \tau') $ for $ \tau-\tau' <0 $
(we have $  G^0=0 $ for this time order).
The first term in the bracket in (\ref{560}) is a typical term for
non-interacting electrons representing that term 
in (\ref{390}) in the first quantized language where the two delta
functions corresponding to the
created and annihilated particle in (\ref{390}) carry the same particle
index. Due to the enormous effort of calculation we determined  the 
terms where the two delta
functions carry different particle indices represented by the rest of the
summands in (\ref{560}) 
only for small distances $ |\vec{x}\,'-\vec{x}| $. 

Because $ G^0 $ or $\vec{A} $, respectively, are  not translation
invariant we obtain
that the Fourier transform of the Green's function
$ G^{\mbox{\scriptsize Weyl}} $ does 
 not have the form
$ G^{\mbox{\scriptsize Weyl}}(\vec{k}, \vec{k}\,';\omega, \omega')=
G^{\mbox{\scriptsize Weyl}}(\vec{k},\omega)
\delta_{\vec{k},-\vec{k}\,'}\delta_{\omega,-\omega\,'} $.  
During the calculation of the Green's function of the
$ \nu=1/\tilde{\phi} $ system 
within CS theories one usually uses the condition 
\begin{equation}
(2 \pi \tilde{\phi})  \rho =
\vec{\nabla}\times  \langle \vec{a} \rangle=\vec{\nabla}\times \vec{A} 
\label{562}
\end{equation}
and
the fact that  $ \vec{A} $ is transversal (Coulomb gauge). 
Moreover, in the
Coulomb gauge we have the freedom to choose the origin
$ (x_0,y_0) $ of the vector
potential $ \vec{A}=1/2 B (y-y_0,-(x-x_0)) $. 
We see from equation (\ref{560}) that the explicit result of the
Green's function depends on the origin of the symmetric gauge potential.
By using the condition (\ref{562})  during perturbational calculations we loose
the dependence of the Green's function \cite{di4} on the external
magnetic field $ \vec{A} $ and thus also on the
origin $ (x_0,y_0) $. Nevertheless, this dependence must appear in through
the limit $ A \to \infty $ in integrals $ \int_{A}d^2r\dots
$ when calculating a Feynman diagram. In Fourier space the dependence of the result on the concrete limit
$ A \to \infty $ should also be seen in the transition going from momentum sums to
momentum integrals. Carrying out this transition  is not straight forward
for integrals of functions 
which fall off very slowly such as the integral of $ f$ (\ref{10}).
The result depends
on the way of calculating the limit $ A \to \infty $.
Because the integrands of perturbational
calculations within RPA fall off rapidly enough  
the explicite dependence on $ (x_0,y_0) $ has to be chosen when going  
beyond RPA.
We see  from (\ref{560}) that there are several ways to choose $ (x_0,y_0) $
or the integral limit $ \int_{A \to \infty} d^2r \dots $, 
respectively, to get a
translationally invariant Green's function (e.g. $x_0=(x_1+x_1')/2$, 
$y_0=(x_2+x_2')/2$, or $x_0=x_1'$, $y_0=x_2'$, or \dots) which is necessary
when defining the effective mass of quasi-particles.
One can see from (\ref{560}) that all these choices differ in their
Green's function by a pure phase factor
$ \exp[i \vec{A}(\vec{r}_{0,2})(\vec{x}-\vec{x}\,')] $
($ \vec{r}_{0,2} $ is a vector not depending on
$ \vec{x}, \vec{x}\,' $ ) which in
  momentum space results 
in different  translations of the momentum variable of the Green's function.  
Thus, we see that all these choices result in the same effective mass. 
It is then clear that by an additional shift of the origin
$ (x_0,y_0) $ of value $- \vec{f}(0) $ 
we can make the gauge factor  
$ \exp[-i \tilde{\phi} \vec{f}(0)(\vec{x}-\vec{x}\,')] $ in (\ref{560})
vanish.
This gauge transformation does not destroy the translational invariance of the
Green's function (\ref{560}) when choosing either of the origins $ (x_0,y_0) $
discussed above. 

Summarizing, in contrast to the RPA result of the Green's
function in the temporal gauge calculated in  section III
the exact Green's function in the temporal gauge is finite.
Nevertheless, we see from (\ref{550}) and (\ref{560}) that the Green's
function has its support on the one-dimensional subspace
$ \vec{x}=\vec{x}\,' $  for system area $ A \to \infty $.
This is due to the infinite degeneracy of the ground state of
non-interacting electrons in a homogeneous magnetic field. More precisely, at
the end of the parallel transport calculated in (\ref{500}) the ground state 
is orthogonal to its phase transformed starting value resulting in the
vanishing of $ \mbox{Ov}(|\vec{x}-\vec{x}\,'|) $ at values
$ |\vec{x}-\vec{x}\,'| \not=0 $. This degeneracy 
of the ground state 
usually is not present in the case of a Coulomb-interacting system.
Now assume that the non-degeneracy of the ground state is present
for all string values and that the ground state energy correction
due to the additional strings is zero. Then we can deduce that the Green's
function in the temporal gauge has no degeneracy having the support in
a one dimensional subspace of the $\vec{x}$, $ \vec{x}\,' $ plane. This is
immediately clear because in this case the overlap function
$ \mbox{Ov}(|\vec{x}-\vec{x}\,'|) $ consists 
of a phase factor for all
$ |\vec{x}-\vec{x}\,'| $ corresponding to a Berry phase \cite{be1}.
This will be shown in the following subsection. 

\subsection{The exact Green's function with Coulomb
  interaction}
It is clear that in contrast to the non-interacting case above one can not
solve exactly the (ground state) eigenvalue problem of $ N $ electrons in a
homogeneous magnetic field in the background  of two separated
magnetic strings in the presence of Coulomb interaction. Nevertheless,
we will show that we get no energy correction due to the two additive
opposite magnetic strings and that when having a non-degenerate ground state
for $ \phi=0 $ the system has a non-degenerate ground state for
all $ \phi $.  In this subsection, we assume the commonly 
believed non-degeneracy of the ground state of the 
interacting half-filled Landau system.
We denote by $ \Psi^n_{\phi,d} $ a normalized
eigen state with energy $ E^n(\phi,d)$ where $ n $ labels the states. 
Furthermore, we
denote the Hamiltonian of the system by $ H^{e,N}_{ss} $ , i.e.
\begin{align}
&  H^{e,N}_{ss}(\phi,d)  =   \sum^N_{i}\frac{1}{2m}
\bigg(-i \vec{\nabla}_i+\vec{A}(\vec{r}_i) +
  \phi\vec{f}(\vec{r}_i+d/2\vec{e}_x)  \nonumber \\
&   -\phi\vec{f}(\vec{r}_i-d/2\vec{e}_x ) \bigg)^2  
 + \frac{1}{2}\sum_{i\not= j} V^{ee}_{\vec{r}_i,\vec{r}_j} \;.
  \label{565}
\end{align}
Here $ N $ is the number of electrons in the $ \nu=1/\tilde{\phi} $ system.  
For simplicity we used a version of the many
particle Hamiltonian in (\ref{565}) where the magnetic field is symmetric
around the origin and the strings are positioned at $ d/2\vec{e}_x $ and
$ -d/2\vec{e}_x $.
With the help of the current operator 
$ \hat{\vec{J}}_{\phi,d}(\vec{r})=1/m \sum_i
\delta(\vec{r}-\vec{r}_i) \left(-i \vec{\nabla}_i+\vec{A}(\vec{r}_i) +
  \phi\vec{f}(\vec{r}_i+d/2\vec{e}_x)
  -\phi\vec{f}(\vec{r}_i-d/2\vec{e}_x )
 \right) $ (the physical current being the real part of the
 expectation value of this current operator) we get 
\begin{equation}
 \partial_{\xi} E^{n}=  
\int \nts{1} d^2r (\partial_{\xi} \vec{F}(\vec{r})) 
\langle \hat{\vec{J}}(\vec{r}) \rangle^n_{\phi,d} \label{570}
\end{equation}
with
\begin{equation}
 \vec{F}_{\phi,d}(\vec{r})=   
\frac{\phi}{m} \left(\vec{f}(\vec{r}+d/2\vec{e}_x)-\vec{f}(\vec{r}-d/2\vec{e}_x
  ) \right) \,,
\label{580}
\end{equation}
where $ \xi $ equals $ d $ or $ \phi $, respectively.
Here we denoted  $ \langle \Psi^n_{\phi,d}|
 \hat{\vec{J}}_{\phi,d}(\vec{r})
|\Psi^n_{\phi,d}\rangle $ by
$ \langle \hat{\vec{J}}(\vec{r}) \rangle^n_{\phi,d} $. 
One can interpret (\ref{570}) as the energy variation that is due to
the induced electrical field resulting from a variation in the string magnetic
field.  
We have the following relation for the eigenfunctions
\begin{align}
&\Psi_{1-\phi,d}^n(\vec{r}_1,.., \vec{r}_N)  \label{590}\\
& 
= \Psi^n_{\phi,d}(-\vec{r}_1,..,-\vec{r}_N)
e^{-i \sum_{i} \nts{2} \mbox{ \scriptsize arg}[\vec{r}_i+d/2\vec{e}_x]}
e^{+i \sum_{i}\nts{2} \mbox{ \scriptsize arg}[\vec{r}_i-d/2\vec{e}_x]}
\nonumber 
\end{align}
with $ E^n(\phi,d)=E^n(1-\phi,d) $ (we denote these related states by the
same label).
Thus, we obtain 
\begin{equation}
\langle \hat{\vec{J}}(\vec{r}) \rangle^n_{1-\phi,d}=
-\langle \hat{\vec{J}}(-\vec{r}) \rangle^n_{\phi,d}\,.   \label{600}
\end{equation}
Summarizing, due to the different signs of the strengths of the two strings 
we obtain an inversion symmetry of the energy spectrum 
with respect to the strength $ \phi=0.5 $. This
inversion symmetry does not depend on the string distance $ d $. In the
following, we calculate the derivate of the energy $ E^n_{\phi,d} $ 
with respect to $ d $. By using (\ref{570}), (\ref{580}) and (\ref{600})
we get
\begin{eqnarray}
 \partial_{d} E^n(\phi,d)
 & = &    \int \nts{1} d^2r \;  (\partial_{d} \vec{F}_{\phi,d}(\vec{r})) 
\langle \hat{\vec{J}}(\vec{r}) \rangle^n_{\phi,d}  \nonumber \\  
&   = &  -\frac{\phi}{1-\phi}
 \int \nts{1} d^2r \, (\partial_{d} \vec{F}_{1-\phi,d}(\vec{r})) 
\langle \hat{\vec{J}}(\vec{r}) \rangle^n_{1-\phi,d} \nonumber \\  
& = &   -\frac{\phi}{1-\phi} \; 
\partial_{d} E^n(1-\phi,d) \,. \label{610}
\end{eqnarray}
Here we used  $ \vec{F}_{\phi,d}(\vec{r})=\vec{F}_{\phi,d}(-\vec{r}) $.
Since $ E^n(\phi,d)=E^n(1-\phi,d) $ we get
\begin{equation}
  \partial_{d} E^n(\phi,d)=0        \label{620}
\end{equation}
for $ \phi \not= 0,1$.
From (\ref{620}) we observe that one should not have any energy correction
due to the two separated magnetic strings. This is in contrast to the exact
derivation of the ground state energy in subsection A where we showed
that one of the ground state energy levels rises in its energy when switching
on the two strings (see Fig. 1). To get the reason for this discrepancy
we have to examine the above derivation a little more carefully. First,
as mentioned in subsection A (below (\ref{398})) the domain of the
Hamiltonian in (\ref{565}) is different for different $ d $´s. Thus, we have
to spread out the magnetic strings (at string width $ R_0 $)
before carrying out the analysis above.
It is then clear that one can deduce (\ref{620}) only when the energy
levels and wave functions are smooth for string width $ R_0 \to 0$.
This should be the case for $ d > 0 $.
It is then clear from
subsection A that some eigenvalues are discontinuous at $ d=0 $ (for $R_0 \to 0
$), i.e. $ \partial_{d} E^{n}(\phi,0) $ is infinite.
This follows from the fact that 
$ \partial_{d} \vec{F}_{\phi,d}(\vec{r})|_{d=0}=(\phi/m)
  (\vec{e}_y/(x^2+y^2) -\vec{f}(\vec{r})2x/(x^2+y^2)) $ scales like
  $ O(1/r^2) $. Thus, we obtain from (\ref{570}) that
$ \partial_{d} E^{n}(\phi,0) $ can be infinite for states
having a non-zero current expectation value. Since we assumed that
the ground state
of the Coulomb interacting system without strings is non-degenerate and
thus $ \langle \hat{\vec{J}}(\vec{r})\rangle_{\phi,0}=0 $ 
we have one energy level which stays invariant when the magnetic
strings are switched on. This level has the same energy as the ground state 
of the system without the strings. Nevertheless,
a level crossing with this state could occur for $ \phi \not=0,1 $.
This should not happen because we see from (\ref{570}) that if we have an
energy discontinuity for some $ \phi$ then we have a discontinuity for all
$ \phi>0 $ and the energy discontinuity is an increasing function of
$ \phi $ (see also Fig. 1).
Then we obtain from the considerations above
and the knowledge that the energy spectrum is in accordance
for $ \phi=0 $ and $ \phi=1$
that we have a non-degenerate ground state for all $ 0 \le \phi \le 1 $
and $ d \ge 0 $. This ground state should 
have the same energy as the ground state of the system without strings.

To get a better physical insight into the fact that the energy of
the ground state remains invariant for $ \phi \not=0 $ in the
following we give a second derivation. In this derivation we use 
the Wigner-von Neumann theorem \cite{wig1}. This theorem
states that for a Hamiltonian which depends on one
parameter one has a level crossing only in the case where  the 
two states which take part in this level
crossing belong to different representations of the symmetry group of the
Hamiltonian. Because of the non-zero overlap of
the ground state for $ \phi=0 $
and $ \phi=1 $ for all $ d  \ge 0 $ and large
system area (by using that the density of the ground state wave function
is homogenous for $ \phi=0 $) we find that these two states belong to the same representation
of the symmetry group of the Hamiltonian. Thus, we see that these
two states are connected
when switching on the flux $ \phi $ (furthermore, during the parallel transport
(\ref{500}) the system remains in this state).  
Now we calculate the second derivative of the energy of this state  
with respect to $ \phi $ for some $ d >0 $. 
This term is proportional
to the magnetic
field correction at the positions of the two strings
that is created by the electrons from the first order current
correction which is itself caused by 
an infintesimal variation $ \delta \phi $ of  $ \phi $ 
(we have a term similar to expression (\ref{270})).
This current correction is due to the acceleration of the electrons
by the induced electric field resulting from the variation $ \delta \phi $.
Because we have no spin degree of freedom we know from Lenz's rule 
that the magnetic field correction tends to reduce the flux variation. 
This results in a non-negative second order
derivate of the energy
with respect to $ \phi $ for all $ \phi $ and $d $. Because we have  
$ \partial_{\phi} E(0,d)=\partial_{\phi} E(1,d)=0 $ for the non-degenerate
ground state we find that the energy of
the state connecting the ground states for $ \phi=0 $ and $ \phi=1 $
remains invariant when switching on the two magnetic strings.

It is clear that the two derivations above that make
predictions on 
the spectrum of the interacting system are not exact proofs
in the sense that they use only the Hamiltonian (\ref{565}) as an input. 
The exact proof is an outstanding problem. 
Nevertheless, both derivations presented above lead to according 
predictions based on meaningful physical input. 

Now we can use the results at the beginning of section IV and subsection B
to obtain the Green's function for the Coulomb interacting system.
We obtain for the Green's function of the CS system in the temporal gauge 
\begin{equation}
G^{\mbox{\scriptsize Weyl},e}(\vec{x}, \vec{x}\,'; \tau, \tau')=
(-1)^{\frac{\tilde{\phi}}{2}}
\mbox{Ov}^e(|\vec{x}-\vec{x}\,'|)
G^e(\vec{x}, \vec{x}\,'; \tau, \tau') . \label{630} 
\end{equation}
Here we neglected a factor
$ e^{-i \tilde{\phi} \vec{f}(0)(\vec{x}-\vec{x}\,')} $
(see the discussion below (\ref{560})). 
$ G^e $ is the many-particle Green's function
for Coulomb interacting electrons in
a Coulomb gauged vector potential $ A $
(the many particle Green's function of the Hamiltonian
$ H_{ss}(0,0) $). We should mention that the expression (\ref{630}) is valid
for all $ \tau $, $ \tau' $. $ \mbox{Ov}^e(|\vec{x}-\vec{x}\,'|) $ is the
overlap function corresponding to $ \mbox{Ov} $ (\ref{550})
in the non-interacting case. As mentioned at the end of subsection C,
$ \mbox{Ov}^e $ is a phase factor which is calculated by the parallel
transport equation (\ref{500}) ($ \Psi_p^\phi $ in this equation has
to be substituted by the non-degenerate ground state of string strength $ \phi $). 
Thus, we see from this equation that the Green's function does not have the
same degeneracy  as in the case of the
non-interacting system. 

\section{Summary and Outlook}
The CS theory of the $\nu=1/2 $ system in the Coulomb gauge
established by HLR does not allow the formulation of a
quasi-particle picture
of the CS fermions without a physical motivated cancellation
of diverging terms in Feynman diagrams.
This is due to a vanishing Green's function for infinite area.
Motivated by this, 
we consider in this paper the CS theory in the temporal gauge.
Our intention is 
to formulate a CS theory of the half-filled Landau level
which has  meaningful quasi-particles. At first, we derive 
the CS path
integral in the temporal gauge by a gauge transformation from the
correct normal ordered CS path integral in  the Coulomb
gauge. We show that one has to be careful with the time slices of the path
integrals to get the correct result. From this we calculate the self energy 
in RPA for both gauges. For the self energy in the temporal gauge, we obtain 
a scaling with $ 1/T $. With the help of a CS
retransformation of the path
integral representing the Green's function we calculate
the Green's function non-perturbatively for the non-interacting electron
system. We get a finite result for this Green's function.
The reason for this misbehaviour  in RPA is a wrong result when calculating
the ground state energy up to  second order perturbation theory in the string
strength for a system of electrons in a homogeneous magnetic field
and two separated magnetic strings of opposite strength.
We obtain exactly a zero  energy correction
of the ground state. Furthermore, we calculate explicitly
the ground state wave functions of this system. By considering
methods also used in
deriving the Berry phase we obtain the exact Green's function in
the temporal gauge.
Apart from this we get the reason for the missing
of the $ \log(A) $ divergence in the self energy in the temporal gauge in RPA.
This is due to  a dynamical creation of 
phase factors linking the created and
annihilated electrons in the Green's function
to all other electrons when switching on adiabatically the two
magnetic strings. These  phase factors
do not exist in the Coulomb gauge thereby leading to the $ \log(A) $
behaviour of the exact Green's function.
As a generalization of these results, we find  that the
ground state energy of an interacting electron system
(in the homogeneous magnetic background
field) should be the same with and without the two strings.
From this we deduce that also the Green's function of the interacting electron
system is finite. 

Summarizing, we showed in this paper that the
CS theory in the
temporal gauge contains a meaningful Green's function 
apart from the physical meaningful dipole feature
of the quasi-particles.
To our understanding, this is the premise to examine the widely
discussed effective mass of the CS fermions. 
The next stage would be the formulation of a perturbation theory of the 
temporal gauge which shows the finiteness of the Green's function.
This should be a conserved approximation \cite{bay1}. The knowledge of the 
exact Green's function of the non-interacting system derived in section IV 
should be useful in finding such a formulation. 
Such a procedure should yield meaningful  
results not only for the 
one particle sector given by the Green's function 
but also new results for the higher 
particle sectors 
 representing most of the physics of the composite fermions. 
This is work in progress. 

\bigskip

We would like to thank W. Weller, M. Bordag and W. Apel for helpful 
discussions during the course of this work. Further we acknowledge
the support of the Graduiertenkolleg ``Quantenfeldtheorie'' at the University
of Leipzig.

\begin{appendix}
  \section{The calculation of the exact $ (a^0,a^0) $ vertex}
In this section we calculate
$ {\cal D}^{\nts{2}\mbox{ \scriptsize  ex}}_{00}(\vec{r},\omega=0) $ (\ref{270}) for a non-interacting electron system 
without any approximation. 
With the help of the eigenfunctions $ \Psi^0_{n,p} $ (\ref{396}) and
$ E^0_{n,p} $ (\ref{398}) we get for the second summand in (\ref{270})
by the insertion of a complete set of
eigenfunctions 
\begin{align}
 &  {\cal D}^{\nts{2}\mbox{ \scriptsize ex},2}_{00}(\vec{r},\omega=0)=\tilde{\phi}^2
 \frac{1}{\sum\limits_{\Psi_{\vec{p}}\in \nu_{1/\tilde{\phi}}} \nts{5} 1}
 \sum\limits_{\Psi_{\vec{p}}\in \nu_{1/\tilde{\phi}}}
\sum\limits_{i=1}^{N}\sum\limits_{n=1}^{\infty}\frac{1}{n \omega_c m^2}
\nonumber \\
 &
\times \nts{1} 2  \mbox{Re} \nts{1}\Bigg[ \nts{2} \int \nts{2} d^2r''
\Psi^{0*}_{0,p_i}(\vec{r}\,'')
\nts{1.5} \left(\nts{1.5} \frac{\vec{\nabla''}}{i}\nts{1} +\nts{1} \vec{A}(\vec{r}\,'')
  \nts{1.5} \right) 
\nts{1.5}\Psi^0_{n,p_i}(\vec{r}\,'')   \vec{f}(\vec{r}-\vec{r}\,'') \nonumber \\
 & \times 
 \int d^2r'\, \Psi^{0*}_{n,p_i}(\vec{r}\,')
\left(\nts{1} \frac{\vec{\nabla'}}{i}+\vec{A}(\vec{r}\,') \nts{1} \right)
\Psi^0_{0,p_i}(\vec{r}\,')   \vec{f}(\vec{r}\,') \Bigg].  \label{a10}
\end{align}
Here $ N $ is the number of electrons in the $ \nu=1/\tilde{\phi} $ system.
$ \mbox{Re} $ is the real part of its argument. 
The subexpression 
$- \sum^\infty_{n=1} 1/(n \omega_c m) \Psi^0_{n,p_i}(\vec{r}\,'')    
 \int d^2r'\, \Psi^0_{n,p_i}(\vec{r}\,')
(\vec{\nabla'}/i+\vec{A}(\vec{r}\,'))
\Psi^0_{0,p_i}(\vec{r}\,')  \vec{f}(\vec{r}\,') ) $  
in (\ref{a10}) is the first order correction to the ground state wave
function $ \Psi^0_{0,p_i} $ 
and is due to a magnetic string at the origin. Thus, we see that $ {\cal
  D}^{\nts{2}\mbox{ \scriptsize ex},2}_{00}(\vec{r},\omega=0) $
corresponds to the first order correction
(in $ \tilde{\phi}' $) of the operator
$ \tilde{\phi} \int \nts{2}  d^2r' \nts{1}
\vec{f}(\vec{r}-\vec{r}\,')\cdot  \vec{J}(\vec{r}\,',t´) $ 
to the perturbation 
$ \tilde{\phi}' \int d^2r'\vec{J}(\vec{r}\,',0 ) 
\cdot \vec{f}(\vec{r}\,') $ (putting at last $\tilde{\phi}'= \tilde{\phi}$).
From (\ref{396}) we obtain  that the
eigenfunctions $ \Psi_{n,p}^{\phi} $ of the one-string system
are analytic in $ \phi $ at $ \phi \approx 0 $ for $ p \not= 0 $.   
Therefore, we can split the n-sum in $ {\cal
  D}^{\nts{2}\mbox{ \scriptsize ex},2}_{00}(\vec{r},\omega=0) $ (\ref{a10})
in terms with $ p_i=0 $ and
terms  with $ p_i \not=0 $. The term in the square brackets in (\ref{a10})
for $ p_i=0 $ can be calculated easily giving zero. Thus, we get for
$ {\cal D }^{\nts{2}\mbox{ \scriptsize ex},2}_{00}(\vec{r},\omega=0) $ 
\begin{align}
& {\cal D}^{\nts{2}\mbox{ \scriptsize ex},2}_{00}(\vec{r},\omega=0) =   
-\frac{\tilde{\phi}^2}{m}
 \frac{1}{\sum\limits_{\Psi_{0,\vec{p}}\in \nu_{1/\tilde{\phi}}} \nts{5} 1}
 \sum\limits_{\Psi_{0,\vec{p}}\in \nu_{1/\tilde{\phi}}}
\sum\limits_{i=1}^{N} (1-\delta_{p_i,0})  \nonumber  \\ 
& 
\qquad  \times \partial_{\tilde{\phi}'} \Bigg[ \nts{2} \int \nts{2} d^2r''
\, \Psi^{\tilde{\phi}'*}_{0,p_i}(\vec{r}\,'')
\left(\nts{1} \frac{\vec{\nabla''}}{i}+\vec{A}(\vec{r}\,'')\right) \nonumber 
\\
  &   
\qquad \times \Psi^{\tilde{\phi}'}_{0,p_i}(\vec{r}\,'')
\vec{f}(\vec{r}-\vec{r}\,'') \Bigg]_{\tilde{\phi}'=0} \,. \label{a20}
\end{align}
This expression as well as the first term in (\ref{270}) 
can be calculated analytically. 
After some straight forward calculations we get (\ref{280}).
  
\section{A monomial basis for  $ HL^2_2 $}
In this appendix we prove that $\{z^p\} $ ($p \in \mathbb{N}_0$) is a basis of $ HL^2_2 $.  
For simplicity we use $ B=1 $ in the following considerations.
Then we have 
\begin{equation}
HL^2_i( \mathbb{C},\alpha_i):=\left\{F \in H( \mathbb{C}) \Big| \int dz |F(z)|^2 \alpha_i(z) <
  \infty\right\} \label{c10}
\end{equation}
where $H( \mathbb{C}) $ are the holomorphic functions of the complex plane $ \mathbb{C} $ and  
\begin{eqnarray}
\alpha_1(z) & = & e^{-\frac{|z|^2}{2}} \,, \label{c20} \\ 
 \alpha_2(z) & = & e^{-\frac{|z|^2}{2}}r^{-2 \phi}\left(r^2-
    2r d \cos{\varphi}+d^2\right)^{\phi}  \,.
    \label{c30} 
\end{eqnarray}
It is shown in \cite{ha1} that $ HL^2_1 $, $ HL^2_2 $ are Hilbert spaces
(with the scalar products (\ref{445}) and (\ref{470})) 
and that the functions $ \{z^p\} $ ($ p \in \mathbb{N}_0 $) form  an
orthogonal  basis
of $ HL^2_1 $. We will
show in the following that $ \{z^p\} $ is also basis of $ HL^2_2 $.

To this end we define
\begin{eqnarray}
s(R) & := & \mbox{sup}_{r \ge R}\left\{\left|r^{-2 \phi}\left(r^2-
    2r d \cos{\varphi}+d^2\right)^{\phi}\right|\right\}\,, \nonumber \\
  \|F\|_{R,i} & := &  \int\limits_{|z| \ge R} dz |F(z)|^2  \alpha_i(z)\,.
\label{c50}
\end{eqnarray}
Here sup is the supremum of the argument.
It holds that $ \| g \|_{R,2} \le s(R)\, \| g \|_{R,1} $ for $ R \ge 0 $.  
Due to the holomorphy of the functions in $ HL^2_1 $, $ HL^2_2 $
we know that  for $ u(z) \in
HL^2_1 $ there exist a series $ \sum_{n \ge 0} a_n z^n =u(z) $ 
of uniform convergence for all finite $ R $ with $ |z| \le R $ and
point wise convergence in the whole complex plane.
Thus, we have $ \lim_{N \to \infty}
(\| \sum_{n \ge 0}^N a_n z^n-g \|_{2}-\| \sum_{n \ge 0}^N a_n z^n-g \|_{R,2})=0
$. Now we have to show that $ \lim_{N \to \infty}\| \sum_{n \ge 0}^N a_n
z^n-g\|_{R,2}=0 $. Then we have the following inequality 
\begin{align}
& \|\sum_{n \ge 0}^N a_n z^n-g\|_{R,2} \le s(R) \,
\|\sum_{n \ge 0}^N a_n z^n-g\|_{R,1}  \nonumber \\
& \le  s(R)\, \|\sum_{n \ge 0}^N a_n z^n-g\|_{1} \,. \label{c60} 
\end{align}
We see from (\ref{c10}), (\ref{c20}) and (\ref{c30}) that 
$ HL^2_1 $ and $ HL^2_2 $ contain the same holomorphic
functions. Because $ \{z^p\} $ is a basis of the Segal-Bargmann space
$ HL^2_1 $ we have from (\ref{c60}) that
$ \lim_{N \to \infty}\| \sum^N_{n \ge 0} a_n
z^n-g\|_{R,2}=0 $ and thus the proof that $ \{z^p\} $ is a basis of 
$ HL^2_2 $.

\section{The asymptotic solutions of the time dependent
  Schr\"odinger equation (\ref{490}) }
In this appendix we discuss the solutions of the
time dependent Schr\"odinger equation (\ref{490}) for $ \beta \to \infty $.
We will show that the solutions of this equation have a vanishing part
in the eigen
subspace corresponding to higher eigenvalues, i.e.   
$ \varphi_A(\phi) $ is zero for $ 0 < \phi \le 1 $. 
First, we get from (\ref{490})
\begin{equation}
   \partial_{\phi} \langle \varphi_A| \varphi_A \rangle
   =- \partial_{\phi} \langle \varphi_0| \varphi_0 \rangle
-2\frac{\beta}{\tilde{\phi}} \langle \varphi_A|
  \left(H^1_{ss}(\phi)-\frac{\omega_c}{2}\right)
    |\varphi_A \rangle .    \label{d10} 
\end{equation} 
Denoting by $ \Delta E(\phi) $ the difference between
the energy of the first excited state and the ground state which is zero 
for $ \phi =0, 1 $  we
define $ \langle \varphi_A'|\varphi_A' \rangle $ by the following
differential equation
\begin{equation}
   \partial_{\phi} \langle \varphi_A'| \varphi_A' \rangle
   =- \partial_{\phi} \langle \varphi_0| \varphi_0 \rangle
-2 \frac{\beta}{\tilde{\phi}}  \Delta E(\phi) \langle \varphi_A'
    |\varphi_A' \rangle \,.    \label{d20} 
\end{equation}
By comparing  (\ref{d10}) and (\ref{d20}) we find that
 $ \langle \varphi_A' |\varphi_A' \rangle \ge
\langle \varphi_A |\varphi_A \rangle $ for $ 0 \le \phi \le 1 $.    
By a straight forward calculation we obtain the following solution of the
differential equation (\ref{d20})
\begin{align}
&  \langle \varphi_A'| \varphi_A' \rangle(\phi)=  
 \langle \varphi_A'| \varphi_A' \rangle(0) \; \exp\left[-2\frac{\beta}{\tilde{\phi}}
   \int\limits_0^{\phi}d\phi' \Delta E(\phi')\right]
  \nonumber   \\
 &-
\int\limits_0^{\phi} d\phi' \, ( \partial_{\phi'}
\langle \varphi_0| \varphi_0 \rangle(\phi')) 
 \exp\left[-2\frac{\beta}{\tilde{\phi}}
   \int\limits_{\phi'}^{\phi}
   d\phi'' \Delta E(\phi'')\right]\,.\label{d30}  
 \end{align}
We see from (\ref{d30}) that
$ \partial_{\phi} \langle \varphi_0| \varphi_0
\rangle(\phi) $ has to be negative and
infinite for some value $ \phi>0  $ 
in order to get a finite value
$ \langle \varphi_A'| \varphi_A' \rangle(\phi)>0  $ for $ \beta \to
\infty $ and $ \phi > 0 $.
In the following we will show that this can not be the case.
From (\ref{490}) we get
\begin{eqnarray}
 \lefteqn{\partial_{\phi}
\langle \varphi_0| \varphi_0 \rangle (\phi)
   =   2 \mbox{Re}\left[\langle \partial_{\phi}
     \varphi_0(\phi)|
  \varphi_A(\phi) \rangle \right]} \label{d40} \\
& = &  2 \mbox{Re}\left[\langle \varphi_0(\phi)|
\log(\sqrt{r^2-2d r \cos{\varphi}+d^2}/r) |\varphi_A(\phi)
 \rangle \right]\,.    \nonumber 
\end{eqnarray}
Here we used the basis functions (\ref{480}).
We see from (\ref{d40}) that
$ \partial_{\phi}
\langle \varphi_0| \varphi_0\rangle (\phi)$ is finite for
$ 0 \le \phi \le 1$. From this, (\ref{d30}) and
the spectrum in Fig. 1 we get that 
$ \langle \varphi_A| \varphi_A \rangle  (\phi) $ vanishes as
$ O( 1/\beta ) $ for $ 0 <\phi<1 $ and as
$ O( 1/\beta^{1/(n+1)} ) $ at $ \phi=1 $ where $ n $ is defined by the
order of the intersection of the eigenvalues at $ \phi=1 $, i.e
$ \Delta E(1-\phi)=O((1-\phi)^n)$. In the case that the
 intersection is smooth we get that 
$ \langle \varphi_A| \varphi_A \rangle  (\phi) $ vanishes as
$ o(1) $ at $ \phi=1 $.
Furthermore, we see from (\ref{d30}) and (\ref{d40})
that the solutions of the Schr\"odinger equation (\ref{490}) are  only non-zero
for starting values $ \varphi_0(0)\not=0 $ in the case  $ \beta \to \infty$.  
\end{appendix}

\end{document}